\documentclass[11pt,a4paper]{article}

\pdfoutput=1
\usepackage{jcappub}
\usepackage{blindtext}
\usepackage{hyperref}
\usepackage{amsmath,amssymb}
\usepackage{float}
\usepackage{microtype}
\usepackage[normalem]{ulem}
\usepackage{graphicx}
\usepackage{bm}
\usepackage{latexsym}
\usepackage{epsfig}
\usepackage{psfrag}
\usepackage{color}
\usepackage[dvipsnames]{xcolor}
\usepackage{subfigure}
\usepackage{graphicx}

\usepackage{amssymb}
\usepackage{amsmath}
\usepackage{bm}
\usepackage{latexsym}
\usepackage{epsfig}
\usepackage{psfrag}
\usepackage{amsmath}
\usepackage[normalem]{ulem}
\usepackage{textcomp}
\usepackage{color}
\usepackage{pstricks}
\usepackage[utf8]{inputenc}
\usepackage{array}
\usepackage{comment}
\usepackage{mathalpha}

\usepackage{tabularray}
\makeatletter
\gdef\@fpheader{}
\makeatother

\def\beq{\begin{equation}}
\def\eeq{\end{equation}}
\def\bea{\begin{eqnarray}}
\def\eea{\end{eqnarray}}
\def\be{\begin{equation}}
\def\ee{\end{equation}}

\def\bse{\begin{subequations}}
\def\ese{\end{subequations}}

\def\te{\tilde{\eta}}
\def\tx{\tilde{x}}

\def\cR{\mathcal{R}}
\def\cP{\mathcal{P}}

\def\cI{\mathcal{I}}

\def\ee{\eta_{\mathrm{e}}}
\def\ere{\eta_{\mathrm{re}}}
\def\eeq{\eta_{\mathrm{eq}}}

\def\ke{k_{\mathrm{e}}}
\def\kre{k_{\mathrm{re}}}
\def\keq{k_{\mathrm{eq}}}
\def\Tre{T_{\mathrm{re}}}
\def\Teq{T_{\mathrm{eq}}}

\def\wre{w_{\mathrm{re}}}
\def\xre{x_{\mathrm{re}}}

\def\vk{\bm{k}}
\def\vp{\bm{p}}
\def\vq{\bm{q}}

\def\HI{H_{_{\mathrm{I}}}}
\def\As{A_{_{\mathrm{S}}}}

\def\ns{n_{_{\mathrm{S}}}}
\def\fnl{f_{_{\mathrm{NL}}}}
\def\Mpl{M_{_\mathrm{Pl}}}

\def\f{\frac}
\def\l{\left}
\def\r{\right}
\def\d{\mathrm{d}}

\def\fpbh{f_{_{\mathrm{PBH}}}}

\def\nn{\nonumber}
\def\wre{w_{\mathrm{re}}}
\def\wf{w_{\mathrm{re}}}
\def\tre{T_{\mathrm{re}}}
\def\teq{T_{\mathrm{eq}}}

\def\kre{k_{\mathrm{re}}}

\def\meq{M_{\mathrm{eq}}}
\def\mbh{M_{_{\mathrm{PBH}}}}

\def\rgw{\rho_{_{\mathrm{GW}}}}
\def\ogw{\Omega_{_{\mathrm{GW}}}}
\def\oR{\Omega_{\mathrm{r}}}

\def\dneff{\Delta N_\mathrm{eff}}

\def\dc{\delta_\mathrm{c}}
\def\dcan{\delta_\mathrm{c}^{\mathrm{an}}}

\def\aend{a_\mathrm{e}}
\def\are{a_\mathrm{re}}
\def\aeq{a_\mathrm{eq}}

\def\kp{k_{\mathrm{peak}}}

\def\nn{\nonumber} 

\def\f{\frac}
\def\l{\left}
\def\r{\right}
\def\d{{\rm d}}
\def\be{\begin{equation}}
\def\ee{\end{equation}} 
\def\bea{\begin{eqnarray}}
\def\eea{\end{eqnarray}}

\def\lsim{\mathrel{\rlap{\lower4pt\hbox{\hskip0.5pt$\sim$}}
 \raise1pt\hbox{$<$}}}         
\def\gsim{\mathrel{\rlap{\lower4pt\hbox{\hskip0.5pt$\sim$}}
 \raise1pt\hbox{$>$}}}         

\allowdisplaybreaks

\begin{document}

\title{Constraining the history of reheating with the NANOGrav 15-year data}
\author{Suvashis Maity$^{1\,\ast}$,}
\emailAdd{$^\ast$suvashis@physics.iitm.ac.in}
\author{Nilanjandev Bhaumik$^{1,2\,\dagger}$,}
\emailAdd{$^\dagger$nilanjandevbhaumik@gmail.com}
\author{Md Riajul Haque$^{1\,\ddagger}$,}
\emailAdd{$^\ddagger$riaj.0009@gmail.com}
\author{Debaprasad Maity$^{3\, \|}$,}
\emailAdd{$^\|$debu@iitg.ac.in}
\author{L. Sriramkumar$^{1\,\mathsection}$}
\emailAdd{$^\mathsection$sriram@physics.iitm.ac.in}
\affiliation{$^1$Centre for Strings, Gravitation and Cosmology,
Department of Physics, Indian Institute of Technology Madras, 
Chennai~600036, India}
\affiliation{$^2$International Centre for Theoretical Physics Asia-Pacific (ICTP-AP),
 University of Chinese Academy of Sciences, Beijing 100190, China}
\affiliation{$^3$Department of Physics, Indian Institute of Technology Guwahati, Guwahati, Assam, India}
\abstract{Over the last few years, primordial black holes (PBHs) have emerged 
as a strong candidate for cold dark matter. 
A significant number of PBHs are produced when the strength of the primordial 
scalar power spectrum is enhanced on small scales (compared to the COBE 
normalized values on large scales).
Such primordial spectra also inevitably lead to strong amplification of the 
scalar-induced, secondary gravitational waves (GWs) at higher frequencies.
The recent detection of the stochastic GW background~(SGWB)
by the pulsar timing arrays~(PTAs) has opened up the possibility of directly 
probing the very early universe. 
Different studies have shown that, when PBHs are assumed to have been formed
during the epoch of radiation domination, the mechanism for the amplification 
of the scalar-induced GWs that is required to explain the PTA data can 
overproduce the PBHs over some ranges of masses. 
In this work, we assume a specific functional form for the primordial scalar
power spectrum and examine the production of PBHs and the scalar-induced secondary 
GWs during the phase of reheating, which precedes the standard epoch of radiation
domination. 
Specifically, we account for the uncertainties in the conditions for the 
formation of PBHs and ensure that the extent of PBHs produced remains within 
the observational bounds.
We find that the scalar-induced SGWB generated during a phase of reheating with 
a steeper equation of state (than that of radiation) fit the NANOGrav 15-year 
data with a stronger Bayesian evidence than the astrophysical scenario involving 
GWs produced by merging supermassive binary black holes.}
\maketitle


\section{Introduction}\label{sec:Introduction}

During the last few decades, the observations of the cosmic microwave 
background~(CMB) have provided the most effective tool to probe the 
early universe prior to the surface of the last scattering.
The increasingly accurate measurements of the anisotropies in the CMB by 
COBE~\cite{Bennett:1996ce,Smoot:1998jt}, WMAP~\cite{WMAP:2008lyn,WMAP:2010qai,
WMAP:2012nax,WMAP:2012fli} and Planck \cite{Planck:2013pxb,Planck:2015fie,
Planck:2018vyg} have defined the era of precision cosmology. 
These observations have led to strong constraints on the parameters that 
characterize the primordial perturbations such as the scalar amplitude~$\As$,
the scalar spectral index~$\ns$, the tensor-to-scalar ratio~$r$, and the scalar
non-Gaussianity parameter~$\fnl$~\cite{Planck:2018jri,Planck:2019kim}.
However, the CMB constrains the primordial perturbations only on large scales
(i.e. wave numbers corresponding to $10^{-4}< k <1\,\mathrm{Mpc}^{-1}$), and 
the corresponding bounds on smaller scales are considerably weaker.

Over the past decade, the detection of gravitational waves (GWs) by the 
Ligo-Virgo-Kagra (LVK) collaboration (in this context, see, for instance, 
Refs.~\cite{LIGOScientific:2016aoc,LIGOScientific:2016dsl,LIGOScientific:2016wyt,
LIGOScientific:2017bnn,LIGOScientific:2017ycc,LIGOScientific:2017vox}) has 
led to a paradigm shift in the approach to primordial cosmology.
It is now widely recognized that GWs provide a unique window to probe the physics
operating in the early universe (see, for example, the recent 
reviews~\cite{Guzzetti:2016mkm,Caprini:2018mtu,Domenech:2021ztg,Roshan:2024qnv}). 
The GWs observed by the LVK detectors were of astrophysical origin, and 
they were produced by the mergers of binary black holes (BHs) or neutron stars.
As we shall discuss below, there exist a variety of phenomena that can generate 
GWs in the primordial universe and, importantly, the GWs produced in the early 
universe will be stochastic in nature.
Moreover, because they interact weakly, GWs carry information about the earliest 
stages of our universe and, hence, they probe primordial physics more effectively 
than the CMB, in particular, on small scales.
But, despite a dedicated effort, the LVK collaboration did not detect a stochastic
GW background (SWGB) over the frequency range of, say, $1<f<10^2\, \mathrm{Hz}$,
that its detectors are sensitive to~\cite{KAGRA:2021kbb}.
However, a few months ago, this barrier was overcome when the Pulsar Timing 
Arrays (PTAs)---such as NANOGrav~\cite{NANOGrav:2023gor, NANOGrav:2023hde}, EPTA
(including the data from InPTA)~\cite{Antoniadis:2023lym,EPTA:2023fyk}, 
PPTA~\cite{Zic:2023gta, Reardon:2023gzh}, and CPTA~\cite{Xu:2023wog}---jointly 
reported the first detection of an SGWB over the frequency range of $10^{-9}< 
f <10^{-6}\, \mathrm{Hz}$.
Specifically, it was found that the SGWB generated by the merging of supermassive 
BH binaries (SMBHBs) do not statistically explain the observed data in an 
efficient manner~\cite{NANOGrav:2023hvm}. 
In fact, the NANOGrav 15-year data favored (with a particularly strong Bayesian 
evidence) primordial origins for the detected SGWB signal~\cite{NANOGrav:2023hvm,
NANOGrav:2023hde,NANOGrav:2023gor}. 
This has immediately fuelled the search for several different channels for the 
cosmological generation of the SGWB.
These include the SGWB from: (i)~models that lead to enhanced formation of 
primordial BHs
(PBHs)~\cite{Inomata:2023zup,Franciolini:2023pbf,Cheung:2023ihl,Balaji:2023ehk,
Firouzjahi:2023lzg,Unal:2023srk,Frosina:2023nxu,Liu:2023ymk,HosseiniMansoori:2023mqh,
Liu:2023pau}, (ii)~domination of ultra-low mass 
PBHs~\cite{Bhaumik:2023wmw}, (iii)~mergers due to 
PBHs~\cite{Depta:2023qst,Gouttenoire:2023nzr}, 
(iv)~phase transitions in the early 
universe~\cite{Salvio:2023ynn,Gouttenoire:2023bqy,Ghosh:2023aum,
An:2023jxf,Jiang:2023qbm,Athron:2023mer,DiBari:2023upq,Li:2023bxy}, 
(v)~collapse of cosmic strings and domain 
walls~\cite{Ellis:2023tsl,Lazarides:2023ksx,Zhang:2023nrs,Yamada:2023thl,Lu:2023mcz, 
Babichev:2023pbf,Ge:2023rce,Li:2023tdx,Kitajima:2023cek,King:2023cgv,Lazarides:2023rqf}, 
(vi)~models of axion inflation~\cite{Unal:2023srk,Niu:2023bsr,Murai:2023gkv}, 
(vii)~blue tilted primary tensor power spectra~\cite{Vagnozzi:2023lwo,Borah:2023sbc, 
Datta:2023vbs,Choudhury:2023kam}, (viii)~primordial magnetic fields~\cite{Maiti:2024nhv} 
and (ix)~other possible scenarios~\cite{Ye:2023xyr,Chen:2024twp,Gangopadhyay:2023qjr}.

In this work, we focus on SGWB generated during the formation of PBHs. 
This is one of the most studied explanations for the PTA data.
Nevertheless, we believe that, in the scenario, there are some caveats, such
as the overproduction of PBHs, which require further investigation.  
PBHs are an attractive candidate for dark matter because of their cold, non-interacting 
nature~\cite{Hawking:1971ei,10.1093/mnras/168.2.399,Carr:2020xqk,Carr:2021bzv}. 
Contrary to astrophysical BHs, which are typically of the order of solar mass, PBHs 
can have masses ranging from a few grams to billions of solar masses. 
PBHs can also play a pivotal role in shaping our understanding of the early universe. 
While they are constrained in a variety of manner over different mass ranges, there still
remains a window over the asteroid mass range where the PBHs can account for the dark 
matter completely~\cite{Montero-Camacho:2019jte}.
There are a number of potential mechanisms through which PBHs can form, including 
bubble collisions~\cite{Hawking:1982ga,Scelfo:2018sny,PhysRevD.50.676}, collapse 
of domain walls~\cite{Rubin:2001yw, Dokuchaev:2004kr, Garriga:2015fdk, Deng:2016vzb,
Ge:2019ihf, Liu:2019lul}, collapse of cosmic strings~\cite{HOGAN198487,HAWKING1989237, 
PhysRevD.43.1106,MacGibbon:1997pu,Planck:2013mgr,Blanco-Pillado:2017rnf}, electroweak 
phase transition~\cite{Huber:2015znp}, first-order phase transitions~\cite{Caprini:2015zlo, 
Dev:2016feu}, and formation of other topological defects~\cite{Figueroa:2012kw,Sanidas:2012ee}. 
However, in this work, we shall be interested in the scenario wherein PBHs are 
formed due to the collapse of the enhanced density perturbations that originated 
during  inflation~\cite{PhysRevD.50.7173,Yokoyama:1998pt,Kinney:2005vj,
Kohri:2007qn,Alabidi:2009bk,Garcia-Bellido:2017mdw,Pattison:2017mbe,
Ballesteros:2017fsr,Hertzberg:2017dkh,Biagetti:2018pjj,Ezquiaga:2018gbw}. 

Inflation corresponds to the earliest phase of accelerated expansion of our universe.
Though it was initially proposed to resolve the horizon problem, the success of 
inflation lies in the fact that it also provides a mechanism for the creation of 
the primordial scalar and tensor perturbations (in this regard, see the 
reviews~\cite{Martin:2003bt,Martin:2004um,Bassett:2005xm,Sriramkumar:2009kg,
Baumann:2008bn,Sriramkumar:2012mik,Martin:2015dha}).
While the inflationary scalar perturbations contribute to the density fluctuations 
and result in the emergence of structure in the universe, the tensor perturbations 
correspond to GWs. 
These perturbations are eventually responsible for the anisotropies observed in 
the CMB and the distribution of the large-scale structure that we see around us 
today.
The observations of the anisotropies in the CMB by missions such as 
Planck~\cite{Planck:2015sxf,Planck:2018jri} and BICEP/Keck~\cite{BICEP:2021xfz} 
allow us to strongly constrain the primordial scalar power spectrum around the 
pivot scale of $k_\ast = 0.05\,\mathrm{Mpc}^{-1}$. 
But, as we mentioned, the constraints on the primordial scalar power spectrum 
are considerably weaker over smaller scales, say, corresponding to wave numbers 
$k \gg 10^5\,\mathrm{Mpc}^{-1}$. 
Due to this reason, the scalar power spectrum can, in principle, be considerably 
amplified on small scales (when compared to the COBE normalized values on large 
scales), leading to the formation of a significant number of PBHs when the 
concerned wave numbers re-enter the Hubble radius after inflation. 
Such power spectra with enhanced power on small scales can be generated during 
inflation with the aid of potentials that contain either a point of 
inflection~\cite{Tsamis:2003px,Kinney:2005vj,
Garcia-Bellido:2017mdw,Ballesteros:2017fsr,Germani:2017bcs,Ezquiaga:2017fvi,
Dalianis:2018frf,Bhaumik:2019tvl,Ragavendra:2020sop} or a small 
bump/dip~\cite{Mishra:2019pzq}.
They can also be generated in inflationary models involving more than one 
field~\cite{Palma:2020ejf,Fumagalli:2020adf,Braglia:2020eai,Pi:2017gih,Wang:2024vfv}, 
due to resonant amplification during inflation~\cite{PhysRevD.102.103527,Peng:2021zon} 
or from squeezed initial states~\cite{Ragavendra:2020vud}. 
At the first order in perturbation theory, the scalar and tensor perturbations
are decoupled and hence evolve independently.
However, at the higher order in the perturbations, one form of perturbation can 
act as a source for the other.
As a result, when the amplitude of the scalar perturbations at the first order is 
enhanced over small scales to produce a significant number of PBHs, they inevitably 
induce tensor perturbations of considerable strengths at the second order.
In such scenarios, the PBHs form almost instantaneously when the wave numbers with
significant scalar power re-enter the Hubble radius and, in the process, also 
generate a SGWB of amplitudes that are comparable to the sensitivities of the 
ongoing and forthcoming GW observatories. 

The formation of PBHs and the generation of scalar-induced, secondary GWs during 
the epoch of radiation domination have been discussed in the literature in good 
detail, particularly in the context of the PTA data. 
The main caveat in this scenario remains the fine-tuning and the possible 
overproduction of PBHs in a certain range of masses~\cite{Inomata:2023zup}. 
This issue has motivated further studies wherein the formation of PBHs has been
considered during non-standard histories post inflation~\cite{Domenech:2020ers, 
Harigaya:2023pmw,Liu:2023pau,Zhao:2023joc,Choudhury:2023fjs,Domenech:2024rks}. 
Reheating is the phase that immediately follows inflation.
During this period, the inflaton oscillates at the bottom of the inflationary 
potential, and, with the aid of suitable couplings, the energy from the inflaton 
is transferred to the fields that constitute the standard model of particle
physics. 
In principle, the PBHs can form during the phase of reheating and, in such a scenario, 
the parameters such as the effective equation-of-state (EoS) during reheating and 
the reheating temperature will affect the formation mechanism as well as the number 
of PBHs produced~\cite{Bhattacharya:2023ztw}. 
An epoch of reheating with a general EoS can also significantly alter the strength 
and the spectral shape of the secondary GWs~\cite{Domenech:2019quo,Domenech:2020kqm,
Domenech:2021ztg}. 
In this study, we focus on the formation of PBHs during the phase of reheating 
and explore the manner in which such a scenario can explain the observed PTA 
data. 
We take into account the uncertainties in the conditions for the collapse of 
the PBHs as well as the existing bounds on the abundance of PBHs in the relevant 
range of masses. 
We assume an analytical expression for the inflationary scalar power spectrum, 
which is in the form of a broken power law.
Such a spectral shape is motivated by the forms of the scalar power spectra that 
typically arise when one considers models of inflation that permit a brief epoch 
of ultra slow roll~\cite{Byrnes:2018txb,Ragavendra:2020sop}. 
We scan our parameter space for an optimal fit to the NANOGrav 15-year data, while 
staying strictly within the observational bounds on the abundance of the PBHs.
We perform a Bayesian analysis and compare our scenario with the possible
astrophysical explanation for the PTA results through the mergers of SMBHBs. 
We find that, when we allow room for uncertainties in the conditions for the 
collapse of PBHs, our model is strongly preferred over the scenario involving 
SMBHBs.

This paper is organized as follows.
In Sec.~\ref{sec:PBHreh}, we shall briefly describe the dynamics of the epoch
of reheating and the formation of PBHs during the epoch. 
In particular, we shall discuss the effects of the parameters describing the 
epoch of reheating on the abundance of PBHs and the fraction of cold dark matter
that can be constituted of PBHs. 
In Sec.~\ref{sec:isgwb_pta} we shall study the associated generation of 
secondary GWs (due to the enhanced scalar power on small scales) and its 
prospects in explaining the observations by the PTAs.
We shall carry out a complete Bayesian analysis when comparing the models with
the NANOGrav 15-year data.
We shall further discuss the implications of our results on the energy scale 
of inflation or, equivalently, on the tensor-to-scalar ratio in the context 
of primary GWs.
In Sec.~\ref{sec:cd}, we shall summarize our results and conclude with a 
discussion on their implications. 

At this stage of our discussion, let us make a few clarifying remarks on 
the conventions and notations that we shall adopt in this work. 
We shall work with natural units such that $\hbar=c=1$, and set the reduced 
Planck mass to be $\Mpl=\l(8\,\pi\, G\r)^{-1/2} \simeq 2.4 \times10^{18}\,
\mathrm{GeV}$.
We shall adopt the signature of the metric to be~$(-,+,+,+)$.
Note that Latin indices shall represent the spatial coordinates, except for~$k$ 
which shall be reserved for denoting the wave number. 
We shall assume the background to be spatially flat
Friedmann-Lema\^itre-Robertson-Walker~(FLRW) 
line element described by the scale factor~$a$ and the Hubble parameter~$H$.
Also, an overdot and an overprime shall denote differentiation with 
respect to the cosmic time~$t$ and the conformal time~$\eta$, respectively.


\section{Formation of PBHs during reheating}\label{sec:PBHreh}

In this section, after a brief overview of the dynamics of reheating, 
we shall describe the formation of PBHs during the epoch.
We shall also compare the extent of PBHs formed with the constraints on the 
quantity~$\fpbh(M)$, which denotes the fraction of PBHs that contribute to 
the energy density of cold dark matter.


\subsection{A brief overview of the dynamics during reheating}\label{sec:reheating}

The phase of reheating connects the epochs of inflation and radiation 
domination. 
Consider the simplest scenario of inflation driven by a single, canonical, 
scalar field, say, $\phi$, governed by the potential~$V(\phi$).
In such a case, after the end of inflation, the inflaton oscillates about 
a minima of the potential.
The inflaton is assumed to be coupled (either directly or indirectly) to 
the fields that constitute the standard model of particle physics. 
As the inflaton oscillates, it is expected to eventually decay into the 
particles that constitute the standard model and thermalize, thereby 
leading to the epoch of radiation domination and setting appropriate 
initial conditions for big bang nucleosynthesis (BBN) to be realized.
Over the years, several reheating mechanisms have been proposed based on
either non-perturbative~\cite{Dolgov:1982th,Kofman:1997yn,Amin:2014eta,
Lozanov:2017hjm,Garcia:2021iag} or perturbative processes with non-gravitational
and gravitational interactions~\cite{Haque:2020zco,Garcia:2020wiy,Clery:2021bwz,
Haque:2022kez,Haque:2023yra}. 
In this work, we shall consider the simple perturbative reheating scenario 
with a constant decay width, say, $\Gamma_{\phi}$, for the inflaton. 
Depending upon the decay width, the process of reheating can be either 
instantaneous with the maximum temperature of $\tre\simeq 10^{15}\, 
\mathrm{GeV}$ or it may be prolonged with the minimum possible temperature 
of $\tre =T_{_{\mathrm{BBN}}} \simeq 4$ MeV~\cite{Kawasaki:2000en,Hannestad:2004px}. 
The duration of the phase of reheating depends on the decay width $\Gamma_{\phi}$ 
and the behavior of the potential~$V(\phi)$ near a minima.

With the introduction of the coupling to the other fields through the decay
width $\Gamma_{\phi}$, the equation of motion describing the inflaton is modified 
to be (in this context, see, for instance, Ref.~\cite{Sriramkumar:2012mik})
\begin{equation}
\ddot{\phi}+(3\,H+\Gamma_\phi)\,\dot{\phi}+V_{\phi}(\phi)=0,\label{eq:EoSphi}
\end{equation}
where, as we mentioned, $H$ is the Hubble parameter and $V_{\phi}=\d V/\d\phi$. 
As the inflaton oscillates about a minima of the potential, we can define 
an effective EoS parameter to describe the phase of reheating as $w_{\phi}
=\langle p_\phi\rangle/\langle\rho_\phi\rangle$, where $\rho_\phi$ and 
$p_\phi$ denote the energy density and pressure associated with the inflaton 
and the angular brackets represent the time average over an oscillation.
If we assume that the potential has the form $V(\phi)\propto \phi^n$ near 
the minimum, upon averaging over a single oscillation, we obtain 
that~\cite{PhysRevD.28.1243}
\begin{equation}
w_{\phi} = \f{n-2}{n+2}.\label{eq:eos-n}
\end{equation}
Note that, for $n=2$, i.e. when the potential behaves quadratically near
the minimum, we have $w_{\phi} = 0$, which indicates that reheating behaves
as a matter-dominated phase. 
Similarly, when $n=4$, i.e. in the quartic case, we have $w_{\phi} =1/3$, 
which corresponds to the EoS during a radiation-dominated epoch. 
In principle, $w_{\phi}$ can be as high as unity, which corresponds to a 
phase of kination, i.e. when the kinetic energy of the scalar field 
completely dominates the potential energy.
In the scenario of perturbative reheating of our interest, the equation of
motion~\eqref{eq:EoSphi} for $\phi$ can be obtained from the following equation 
that governs the energy density~$\rho_\phi$ of the scalar field:
\begin{equation}
\dot\rho_{\phi}+ 3\,H\,(1+w_{\phi})\,\rho_{\phi} 
= -\Gamma_{\phi}\,  \,(1+w_{\phi})\,\rho_{\phi}.
\end{equation}
The conservation of the total energy of the system involving the scalar 
field and radiation then leads to the following equation for the energy
density of radiation~$\rho_{_\mathrm{R}}$:
\begin{equation}
\dot \rho_{_{\mathrm{R}}}+ 4\,H\, \rho_{_{\mathrm{R}}} 
= \Gamma_{\phi}\,  (1+w_{\phi})\,\rho_{\phi}.
\end{equation}
Due to the transfer of energy from the inflaton, the dimensionless energy
density of radiation, viz. $\rho_{_{\mathrm{R}}}/(\rho_{_{\mathrm{R}}}+\rho_{\phi})$,
keeps growing until it approaches unity, marking the end of reheating at a
given conformal time, say, $\ere$. 
It is often found that, throughout the entire period, except towards the end,
the reheating phase is governed by the inflaton EoS~$w_\phi$.
Hence, we shall assume that the EoS is constant throughout the reheating phase 
at a given value~$\wre =w_\phi$.
This allows us to describe the reheating phase in terms of only two 
parameters:~the EoS parameter during reheating~$\wre$ and the reheating temperature~$\tre$, 
which is the temperature of the radiation when reheating is complete.

During the phase of reheating with a general EoS parameter~$\wre$, upon using 
the first Friedmann equation and the equation governing the conservation of energy, 
we can arrive at the following expressions for the scale factor and the Hubble 
parameter:
\begin{equation}
a(\eta) \propto \eta^{2/(1+3\, \wre)},\quad 
H(\eta) \propto \eta^{-3\,(1+\wre)/(1+3\,\wre)}\,.\label{eq:w-e}
\end{equation}    
On matching the scale factor and the Hubble parameter at $\ere$, i.e. at the 
conformal time at the end of reheating, we can arrive at the corresponding 
expressions for the scale factor~$a$ and the Hubble parameter~$H$ during 
the epoch of radiation domination to be
\begin{eqnarray}
a(\eta) \propto \eta -\l(\f{1-3\,\wre}{2}\r)\,\ere,\quad
H(\eta) \propto \l[\eta -\l(\f{1-3\,\wre}{2}\r)\,\ere\r]^{-2}.\label{eq:rd}
\end{eqnarray}
Let $\kre=a(\ere)\,H(\ere)$ and $\keq=a(\eeq)\,H(\eeq)$ be the wave numbers that 
re-enter the Hubble radius at the end of reheating and at radiation-matter equality
(corresponding to conformal time~$\eeq$), respectively.
Using the conservation of entropy, we can relate~$\kre$ to~$\keq$ as follows:
\begin{equation}
\f{\keq}{\kre} =\f{\are}{\aeq}
=\l(\frac{g_{\mathrm{s},\mathrm{eq}}}{g_{\mathrm{s},\mathrm{re}}}\r)^{1/3}\,
\f{\Teq}{\Tre},\label{a-r0}
\end{equation}
where, it can be shown that, $\aeq^{-1}=24000\, \Omega_\mathrm{m}\, h^2$, 
$\keq =0.07\,\Omega_\mathrm{m}\, h^2 \,\mathrm{Mpc}^{-1}$ and the temperature 
at equality being $\teq= 5.6\,(\Omega_\mathrm{m}\, h^2)\, \mathrm{eV}$. 
Note that we have expressed all the expressions in terms of the present 
day matter density parameter $\Omega_\mathrm{m}\, h^2=0.13$~\cite{Durrer:2008eom}.


\subsection{Effects of reheating on the formation of PBHs}\label{secPBH}

In this section, we shall discuss the formation of PBHs due to the enhancement in
power on small scales in the spectrum of curvature perturbations generated during 
inflation. 
In the simple picture wherein reheating occurs instantaneously after inflation,
all the small-scale modes re-enter the Hubble radius during the epoch of radiation 
domination.
When scalar perturbations with significant amplitudes re-enter the Hubble radius,
the overdense region collapses almost instantaneously to form PBHs.
The mass of the PBHs thus produced turns out to be nearly equivalent to the mass 
within the Hubble radius at the time of re-entry. 
When there occurs a phase of reheating, some of the small-scale modes re-enter the
Hubble radius during reheating. 
Even in such a situation, as the reheating dynamics can be described by an 
effective EoS, the mechanism for the formation of the PBHs proves to be the 
same. 
However, apart from the amplitude of the scalar power spectrum, the efficiency 
of the formation of PBHs during reheating depends on the EoS.
We shall focus on the collapse of the overdense regions due to the large scalar 
perturbations generated during inflation, which ensures instantaneous collapse. 
We should mention that the location of the peak in the inflationary scalar power 
spectrum determines the time of re-entry of that scale, and thus the time of PBH 
formation. 
Late-forming PBHs correspond to peaks at smaller wave numbers and thus lead to 
PBHs of higher masses. 
PBHs with mass $M < 10^{15}\, \mathrm{g}$ would have evaporated by today due 
to Hawking radiation. 
In this work, we shall be interested in PBHs with mass $M > 10^{15}\,
\mathrm{g}$, which can contribute to the cold dark matter. 

In our analysis, we shall not consider any specific inflationary potential 
for generating the scalar power spectrum.
Rather, we shall consider an analytical scalar power spectrum with a broken 
power law form.
On large scales, we shall assume that the power spectrum is nearly scale 
invariant, as is required to fit the CMB data.
On small scales, we shall assume that the power spectrum rises as $k^4$ as 
it approaches the peak and, beyond the peak, it falls as $k^{n_0}$ with $n_0
< 0$.
We should point out that such power spectra arise in single-field models of 
inflation that permit a brief period of ultra slow roll~\cite{Byrnes:2018txb}. 
The complete form of the scalar power spectrum that we shall work with is given 
by
\begin{align}
\cP_\cR(k)=\As \l(\f{k}{k_\ast}\r)^{\ns-1} 
+A_{0}\l\{\begin{aligned}
&\l(\f{k}{\kp}\r)^{4}\hskip 0.1em &k\leq \kp,\\
&\l(\f{k}{\kp}\r)^{n_0}\hskip0.1em & k\geq \kp,
\end{aligned}\r.\label{eq:PR}
\end{align}
where $\As$ and $\ns$ are the amplitude and spectral index of the power
spectrum at the CMB pivot scale of $k_\ast=0.05\,\mathrm{Mpc}^{-1}$. 
Note that $\kp$ denotes the wave number at which the peak is located and 
$A_0$ represents the extent of enhancement of the power spectrum at the
location of the peak when compared to the nearly scale-invariant amplitude 
on large scales.
Later, when we calculate the extent of PBHs produced and the strengths of
the secondary GWs generated, we shall assume the values of $\As$ and $\ns$ 
to be as suggested by the recent Planck data~\cite{Planck:2018jri}.

To calculate the abundance of PBHs, the two most popular methods that are often
used in the literature are the Press-Schechter formalism~\cite{Press:1973iz} and
peak theory~\cite{Bardeen:1985tr}.  
In general, the results from these two established methods for calculating the
probability of collapse to form PBHs do not converge, even when all the other
conditions and parameters are fixed (in this context, see Refs.~\cite{Musco:2018rwt,
Yoo:2018kvb,Young:2019osy, Bhaumik:2019tvl}). 
Also, there have been recent efforts from different groups to understand the 
efficiency of collapse leading to the formation of PBHs with the aid of more 
involved and accurate numerical frameworks using the so-called compaction function
methods (see, for example, Refs.~\cite{Escriva:2019phb,Musco:2020jjb, Gow:2020bzo,
Harada:2023ffo,Musco:2023dak,Germani:2023ojx,Harada:2024trx}).
These efforts indicate that small differences in the details of implementation
lead to rather different final results. 
In this work, we shall follow the simple Press-Schechter formalism which considers
a Gaussian random field for the initial density fluctuations and a constant critical
density contrast for the formation of PBHs.

Let $P(\delta)$ denote the probability distribution of the density 
contrast~$\delta$, and let $\delta_\mathrm{c}$ denote the critical value of 
the density contrast above which the overdense regions collapse to form PBHs. 
The fraction of the density that collapses to form PBHs (often referred to 
as the PBH mass fraction), i.e. $\beta(M)= \rho_{_{\mathrm{PBH}}}
/{\rho_{\mathrm{total}}}$, is then given by
\begin{equation}
\beta(M) = \int_{\dc}^\infty \d\delta\, P(\delta) 
\simeq \f{1}{2}\,\mathrm{erfc}\l\{\f{\dc}{\sqrt{2}\, 
\sigma_\delta(M)}\r\},\label{eq:beta}
\end{equation}
where, in arriving at the final expression, we have assumed that the 
probability distribution~$P(\delta)$ is a Gaussian function.
During the phase of reheating or the epoch of radiation domination, the 
power spectrum, say, $\cP_\delta(k)$, associated with the density contrast 
is related to the primordial power spectrum of the curvature perturbation 
$\cP_\cR(k)$ through the relation~\cite{Wands:2000dp}
\begin{equation}
\cP_\delta(k)=\l[\f{2\,(1+\wf)}{5+3\,\wf}\r]^2\, \l(\f{k}{a\,H}\r)^4\,\cP_\cR(k).
\label{eq:curv-matter}
\end{equation} 
If $\sigma_\delta(R)$ represents the variance in the density fluctuations
in space, which is arrived at by smoothing over a radius $R$ with the 
help of the window function $W(k\,R)$, then the quantity is given in 
terms of the power spectrum $\cP_\delta(k)$ through the relation
\begin{equation}
\sigma_{\delta}^2(R) = \int \d\ln k\,\mathcal{P}_{\delta}(k)\, W^2(k\,R).
\label{sigg}
\end{equation}
As is often done, we shall assume the window function to be of the 
following Gaussian form:~$W(k\,R) = \exp\,(-{k^2\,R^2}/{2})$.
We are yet to relate the radius~$R$ and the mass~$M$ that the variance 
$\sigma_\delta$ depends on.
We shall discuss this point below.

Let $M_{_{\mathrm H}}$ be the total mass within the Hubble radius $H^{-1}$
at a given time, and let $\meq$ be the mass within the Hubble radius at the 
time of radiation-matter equality.
The mass $M$ of the PBHs formed due to collapse is related to the
quantity $M_{_{\mathrm H}}$ at the time of formation as 
follows~\cite{Carr:1974nx}:
\begin{equation}
M \equiv \gamma\, M_{_{\mathrm{H}}} 
=\gamma\, \f{4\,\pi\,\rho_{\mathrm{total}}}{3\,H^{3}}\biggm \vert_{k=a\,H}, 
\end{equation}
where $\rho_{\rm total}$ denotes the total energy density of the background 
and $\gamma$ is the efficiency of collapse given 
by~$\gamma= \wf^{3/2}$~\footnote{At this point, we should mention 
that there exists a more accurate method to estimate the mass~$M$ of the PBHs, with 
the mass being sensitive to the EoS parameter, say, $\wre$, of the cosmological 
background fluid as well as the profile of the scalar fluctuations~\cite{Musco:2012au,
Musco:2008hv,Hawke:2002rf,Niemeyer:1997mt,Escriva:2021pmf,Escriva:2019nsa,
Escriva:2020tak,Escriva:2021aeh}.
In the study of critical phenomena in gravitational collapse, numerical simulations
have demonstrated that the mass of PBHs obey a scaling relation when the amplitude 
of the perturbation is close to the critical value (i.e. when $\delta_{\rm m} 
- \delta_{\rm c} \leq 10^{-2}$, where $\delta_{\rm m}$ is the initial amplitude of the
perturbation).
According to the scaling relation, the mass of the PBHs is given by
\begin{align}
M=\kappa\,M_{_{\rm H}}\,\l(\delta_{\rm m}-\dc\r)^{\gamma_{\rm f}},
\end{align}
where the exponent $\gamma_{\rm f}$ depends on the EoS parameter~$\wre$, while the 
overall constant~$\kappa$ and the critical density~$\dc$ depend both on~$\wre$ and 
the shape of the scalar power spectrum (in this context, see Refs.~\cite{Evans:1994pj,
Musco:2012au,Escriva:2021pmf}; for a discussion on the dependence of $\gamma_{\rm f}$
on $\wre$, see Ref.\cite{Maison:1995cc}, Tab.~1 and Ref.~\cite{Escriva:2022duf}, Fig.~7).
However, we find that analytical estimates of~$\gamma_\mathrm{f}$ and~$\kappa$, which
would have been helpful, are yet to be arrived at in the literature.}.
For a given reheating temperature $\tre$, we can relate the mass $M$ of the
PBHs at the time of their formation to the quantity $\meq$ as
\begin{equation}
\f{M}{\meq} 
= \gamma\,\l(\f{g_{\mathrm{eq}}}{g_{\mathrm{re}}}\r)^{1/2}\,
\l[ \l(\f{43}{11\,g_{\mathrm{s},\mathrm{re}}}\r)^{1/3}\,
\sqrt{\f{\pi^2\, g_{\mathrm{re}}}{90}}\,
\f{T_0}{k}\r]^{\f{3\,(1+\wf)}{1+3\,\wf}}  
\l(\f{\teq}{\Mpl}\r)^2 \l({\f{\tre}{\Mpl}}\r)^{\f{1-3\,\wf}{1+3\,\wf}},
\label{eq:mass-tre}
\end{equation}
where $g_{\mathrm{re}}$ and  $g_{\mathrm{eq}}$ are the effective relativistic 
degrees of freedom that contribute to the energy density of radiation at the 
times of reheating and equality, respectively, and $g_{\mathrm{s},\mathrm{re}}$ 
represents the number of relativistic degrees of freedom that contribute to 
the entropy of radiation at the end of reheating. Moreover, $T_0$ represents the 
temperature of the CMB today.
We now need to relate the wave number~$k$ that appears in the above expression
for~$M$ to the radius~$R$ over which the variance $\sigma_\delta$ is smoothed
over.
In the absence of any other scale in the problem, we shall assume that $k=R^{-1}$,
which then allows us to relate the radius $R$ to the mass $M$ of the PBHs.

The quantity $\beta(M)$ is exponentially sensitive to~$\dc$ and there is a 
significant amount of uncertainty in choosing a suitable value for~$\delta_c$ 
(in this regard, see, for instance, Ref.~\cite{Bhaumik:2019tvl},  App.~B). 
The critical value of the density contrast $\dc$ should depend on the shape of 
the curvature power spectrum~\cite{Germani:2018jgr,Musco:2020jjb}. 
Also, even for a Gaussian distribution of scalar perturbations, the non-linear
relationship between the density contrast and the curvature perturbation leads 
to a non-Gaussian distribution for the density contrast~\cite{Young:2019yug,
DeLuca:2019qsy}.
Moreover, during inflation, the mechanism that produces the scalar power 
spectrum with sharp features can also generate large levels of 
non-Gaussianities~\cite{Hazra:2012yn,Ragavendra:2020sop,Ragavendra:2020vud}.
Further, many models of PBH formation predict a non-Gaussian tail for the
primordial distribution of the curvature perturbation (in this context, see, 
for example, Refs.~\cite{Figueroa:2020jkf,Cai:2022erk}). 
Since, in this work, we do not consider any particular model of inflation, 
we shall leave a more accurate estimation of the combined effects of 
non-Gaussianities, arising during inflation and due to the non-linearities 
post-inflation, for future work.
We shall assume that the threshold value of the density contrast is given
by the following analytical expression~\cite{Harada:2013epa}:
\begin{equation}
\dc^\mathrm{an}
=\f{3\,(1+\wf)}{5+3\,\wf}\,\sin^2\l(\f{\pi\,\sqrt{\wf}}{1+3\,\wf}\r).\label{eq:cd}
\end{equation}
It is important to recognize that different models for the collapse of PBHs lead 
to different values of~$\dc$.
In what follows, we shall utilize the above analytical estimate for~$\dc$ to 
investigate the dependence of the number of PBHs formed on the reheating 
parameters. 
In due course, we shall also discuss the effects of the variation of~$\dc$. 

Let $\fpbh(M)$ denote the present-day, fractional contribution of PBHs to the
density of cold dark matter, i.e. $\fpbh = \Omega_{_{\mathrm{PBH}}}/\Omega_{\mathrm{c}}$.
Note that the energy density of PBHs~$\rho_{_{\mathrm{PBH}}}$ always varies 
as $a^{-3}$ whereas, during reheating, the energy density of the background 
behaves as $\rho_\mathrm{total} \propto a^{-3\,(1+\wf)}$.
Therefore, during reheating, $\rho_{_{\mathrm{PBH}}}/\rho_\mathrm{total} 
\propto a^{3\,\wf}$. 
In contrast, during the epoch of radiation domination which follows reheating, 
since total energy density varies as $a^{-4}$, we find that $\rho_{_{\mathrm{PBH}}}/
\rho_\mathrm{total} \propto a$. 
Thus, one can express $\fpbh(M)$ corresponding to mode $k$ as
\begin{equation}
\fpbh(k) = \f{\beta(k)}{0.42}\,
\l(\f{k}{\kre}\r)^{\f{6\,\wf}{1+3\,\wf}}\, \l(\f{\aeq}{\are}\r)
=\f{\beta(M)}{0.42} \l(\f{k}{\kre}\r)^{\f{6\,\wf}{1+3\,\wf}}\,
\l(\f{g_{\mathrm{s},\mathrm{re}}}{g_{\mathrm{s},\mathrm{eq}}}\r)^{1/3}\,
\f{\tre}{\teq},\label{fpbh}
\end{equation}
where we have set $\Omega_{\mathrm{c}}^{\mathrm{eq}}=0.42$. 
From Eqs.~\eqref{eq:mass-tre} and~\eqref{fpbh}, we can see that both the mass of
PBHs and their abundance depend on~$\tre$ and $\wre$. 
After some algebra, it can be shown that the quantity $\fpbh(M)$ can be expressed 
as follows:
\begin{equation}
\fpbh(M) = \beta(M)\,
\f{\Omega_{\mathrm{m}}\,h^2}{\Omega_{\mathrm{c}}\,h^2}\,
\l(\f{g_{\mathrm{s},\mathrm{eq}}}{g_{\mathrm{s},\mathrm{re}}}\r)\,
\l(\f{g_{\mathrm{re}}}{g_{\mathrm{eq}}}\r)^{\f{1}{1+\wre}}\,
\l(\f{\tre}{\teq}\r)^{\f{1-3\,\wre}{1+\wre}}\,
\l( \f{M}{\gamma\,\meq} \r)^{-\f{2\,\wre}{1+\wre}}.\label{eq:fpbh}
\end{equation}

\begin{figure}[!t]
\includegraphics[width=0.329\linewidth]{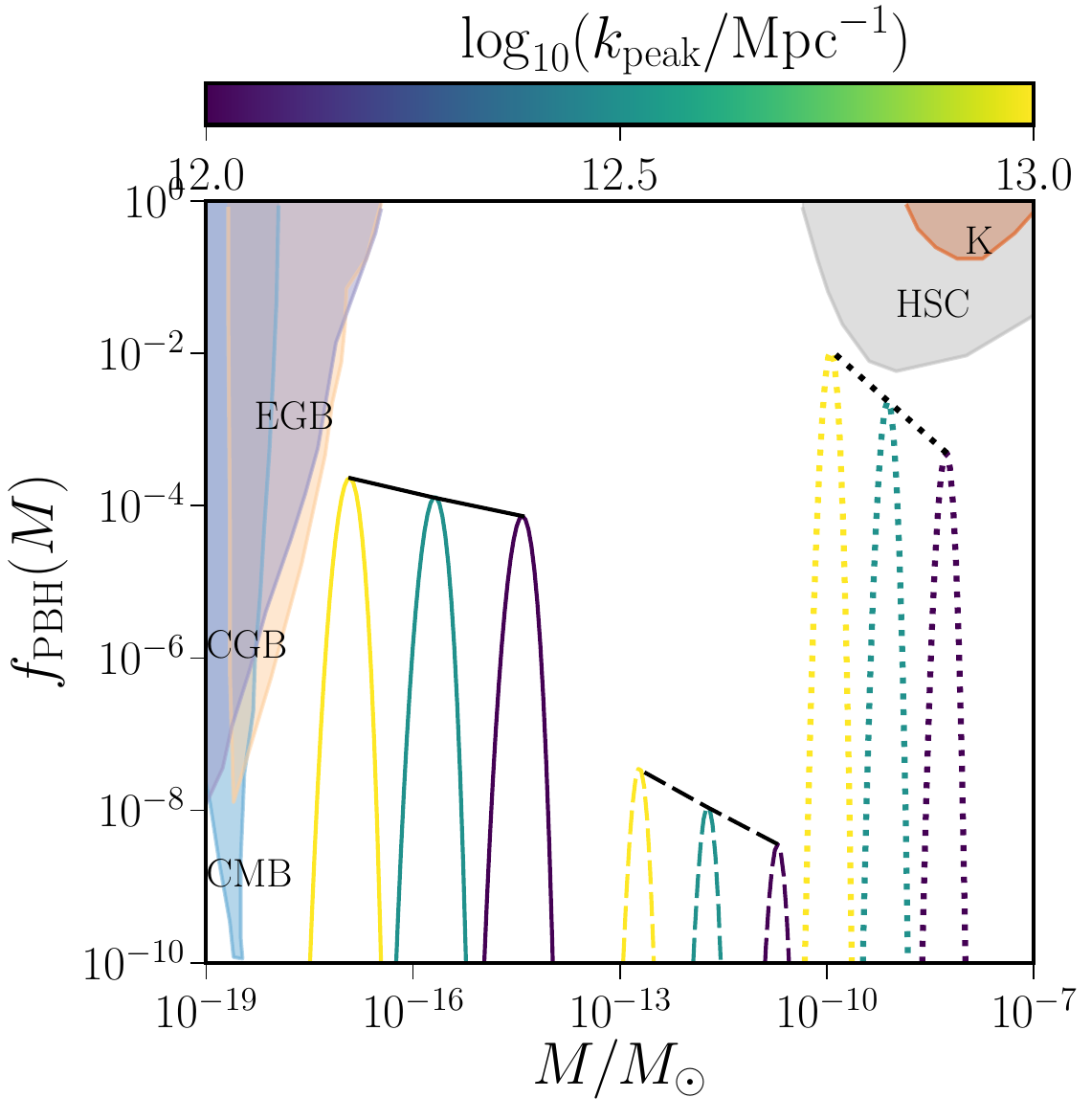}
\includegraphics[width=0.329\linewidth]{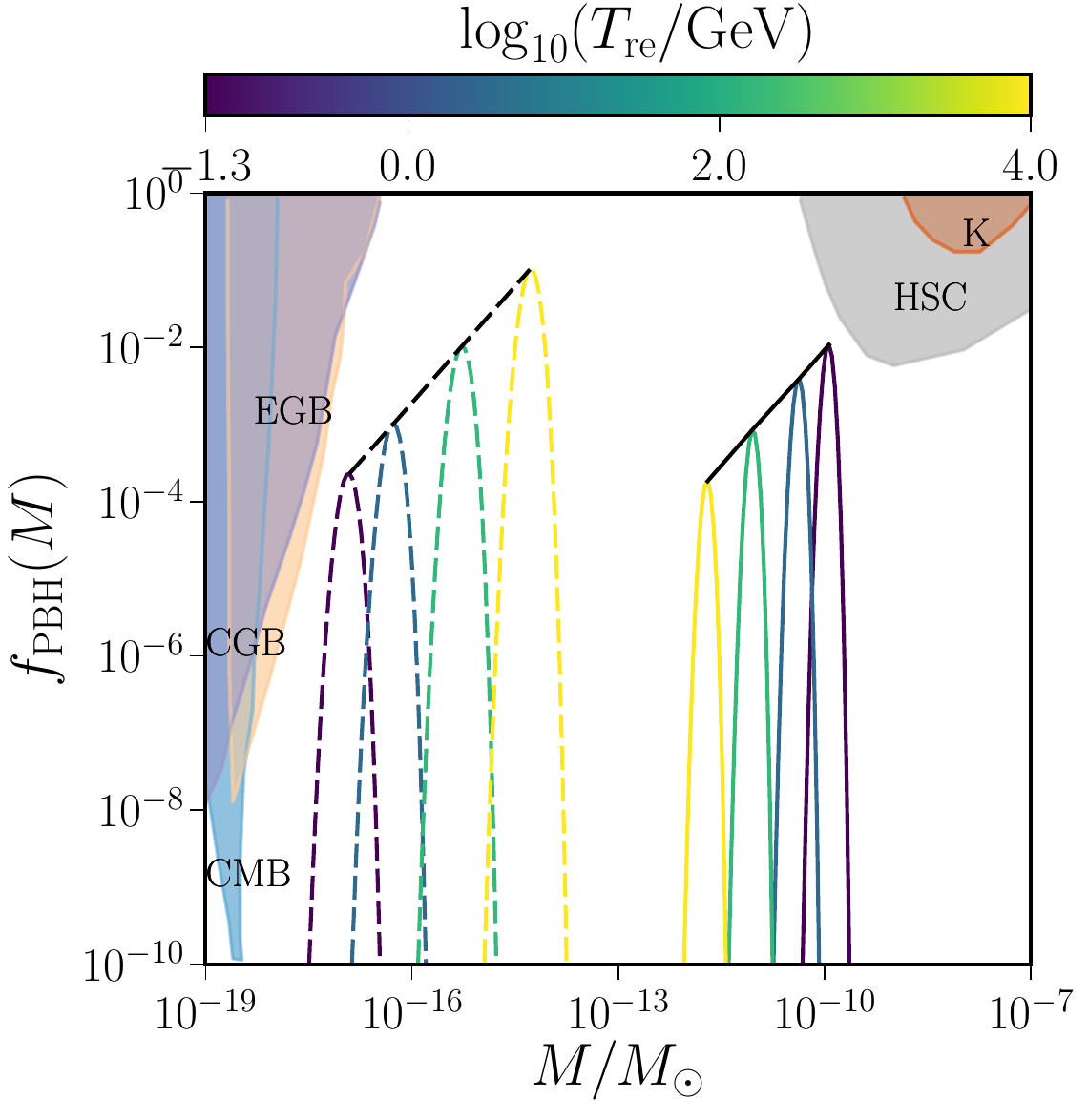}
\includegraphics[width=0.329\linewidth]{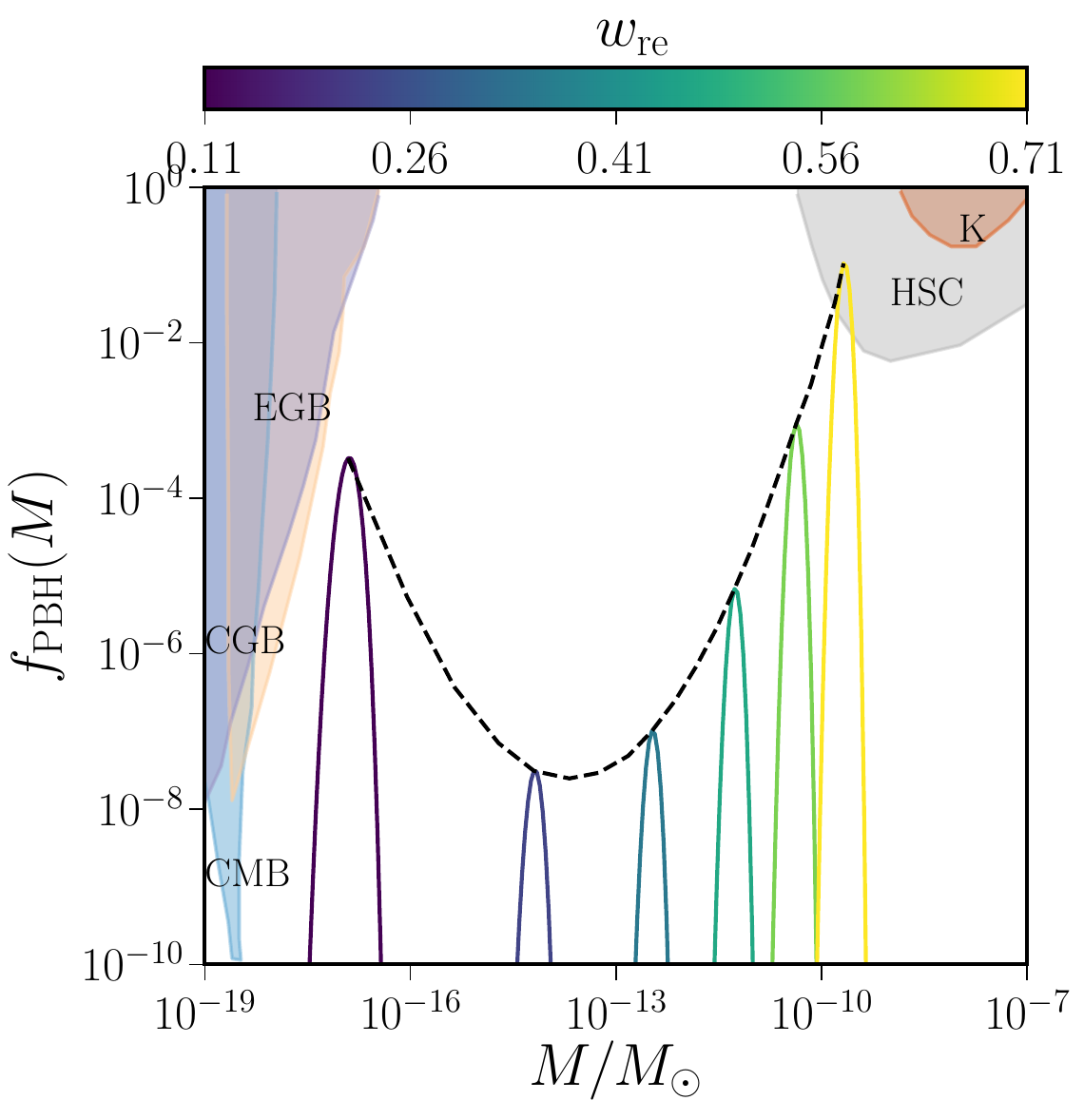}
\caption{The quantity $\fpbh$ is plotted as a function of $M/M_\odot$ 
for a range of $\kp$, $\tre$ and $\wre$ (in the left, middle and right
panels).
We have set $A_0=10^{-1.5}$ and $n_0=-2$ in plotting these figures. 
In the left panel, we have fixed $\tre=50\, \mathrm{MeV}$ and have 
plotted $\fpbh(M)$ for three different values of $\kp$ and for three 
different values of $\wre$, viz. $\wre= (1/9,1/3,2/3)$ (as solid, dashed 
and dotted lines).
We should mention that the peak values of $\fpbh(M)$ behaves as $M^{-2\,
\wre/(1+\wre)}$ and, in particular, as $M^{-1/2}$ when 
$\wre=1/3$~\cite{Ragavendra:2023ret}.
In the middle panel, we have fixed $\kp$ at $10^{13}\,\mathrm{Mpc}^{-1}$ 
and have plotted $\fpbh(M)$ for four different values of $\tre$ and for 
the two different values of $\wre=(1/9,2/3)$ (as solid and dashed lines). 
The slopes of the peaks of $\fpbh(M)$ in this case behave approximately
as $\pm 1$. 
In the right panel, we have set $\tre=50\, \mathrm{MeV}$ and $\kp=10^{13}\,
\mathrm{Mpc}^{-1}$ and have plotted $\fpbh(M)$ for the six different values 
of $\wre$ in the range $0.1 < \wre < 0.7$. 
We see that $\fpbh(M)$ has the lowest value when $\wre=1/3$ and it keeps 
increasing for values away from it.}\label{fig:fpbh}
\end{figure}

In Fig.~\ref{fig:fpbh}, assuming specific values of the parameters $A_0$ 
and $n_0$ that describe the inflationary scalar power spectrum, we have
illustrated the behavior of $\fpbh(M)$ for a range of values of the 
parameters $\kp$, $\tre$ and $\wf$.
In the left panel of the figure, assuming $\tre=50\, \mathrm{MeV}$, we have
plotted the quantity $\fpbh(M)$ for values of $\kp$ that lie in the range 
$10^{12}<\kp< 10^{13}\,\mathrm{Mpc}^{-1}$. 
In the panel, we have also plotted the quantity for three values of $\wre$, 
viz.~ $\wre=(1/9,1/3,2/3)$.
We find that the corresponding slopes of the peaks of $\fpbh(M)$ behave as 
$-1/5$, $-1/2$, and $-4/5$, as suggested by Eq.~\eqref{eq:fpbh}.
In the middle panel, assuming $\kp=10^{12}\,\mathrm{Mpc}^{-1}$, we have 
illustrated the behavior of $\fpbh(M)$ for different values of $\tre$,
with $\wre$ set to $1/9$ and $2/3$.
On combining Eqs.~\eqref{eq:mass-tre} and~\eqref{eq:fpbh}, we can see that, 
as $\tre$ is changed, for $\wre<1/3$, $\fpbh(M)$ behaves as $M$, whereas, for 
$\wre>1/3$, it varies as $M^{-1}$, a point which is also evident from the 
middle panel.
(Though, we should clarify that there is a slight deviation from such behavior 
due to the temperature-dependent coefficients $g_{\mathrm{re}}$ and $g_{\mathrm{s},
\mathrm{re}}$.)
In the right panel of the figure, assuming $\kp=10^{13}\,\mathrm{Mpc}^{-1}$ and 
$\tre=50\, \mathrm{MeV}$, we have plotted $\fpbh(M)$ for six different values of 
$\wre$, ranging from $0.1$ to $0.7$. 
It is clear that the maximum value of $\fpbh(M)$ is the lowest when $\wre=1/3$ 
and it increases as we move away from $\wre=1/3$.
In the figure, apart from $\fpbh(M)$ we have calculated, we have indicated the 
current observational constraints on the quantity. 
The constraints that we have included are from the extragalactic gamma-ray 
background (EGB)~\cite{Carr:2009jm,Arbey:2019vqx}, from the CMB~\cite{Poulin:2016anj, 
Clark:2016nst}, 
and from the microlensing searches by Subaru~(HSC)~\cite{Niikura:2017zjd,Croon:2020ouk} 
as well as Kepler~(K)~\cite{Griest:2013aaa}.
We should clarify that these constraints have been arrived at assuming a 
monochromatic distribution of the PBH mass and they need to be modified 
for extended mass functions.

It is evident from our analysis that the mass $M$ of the PBHs and the 
quantity $\fpbh(M)$ not only depend on the inflationary scalar power 
spectrum but also on the parameters describing the epoch of reheating.
However, certain behavior, 
such as $\fpbh(M) \propto {\Tre}^{\f{1-3\,\wre}{1+\wre}}\,{M}^{-\f{2\,
\wre}{1+\wre}}$ remains valid irrespective of the shape of the power 
spectrum, whereas the actual value of the PBH mass fraction naturally 
depends on all the parameters involved. 
We should also point out that, for the power spectrum~\eqref{eq:PR} we have 
considered, the $\fpbh(M)$ proves to be nearly monochromatic in nature.

There is another point that we need to clarify at this stage of our discussion.
Note that the pre-factor~$(1+\wre)/(5+3\,\wre)$ in $\cP_\delta(k)$ and $\dcan$
[see Eqs.~\eqref{eq:curv-matter} and~\eqref{eq:cd}], which originates from the 
Poisson equation, cancel each other on substituting them in the 
expression~\eqref{eq:beta} for~$\beta(M)$.
We find that the term $\sin^2[\pi\,\sqrt{\wf}/(1+3\,\wf)]$ that appears in the
expression~\eqref{eq:cd} for $\dcan$ has a global maximum at $\wre=1/3$.
Since the maximum of $\dcan$ occurs at $\wre=1/3$, the abundance of PBHs has a 
minimum at the value.
It is due to this reason that, at the time of radiation-matter equality, the 
quantity $\fpbh(M)$ exhibits a similar dependence~on $\wre$, as we have shown 
in the right panel of Fig.~\ref{fig:fpbh}.
However, we should add that such a dependence of $\beta(M)$ on $\wre$ follows from
the formalism we have adopted to estimate the number of PBHs formed~\cite{Harada:2013epa}.
Such a feature may not arise if one follows a different approach to calculate the 
abundance of PBHs (in this regard, see, for example, Ref.~\cite{Domenech:2024rks}).

In the following section, we shall analyze the imprints of the inflationary 
scalar power spectrum and the dynamics of reheating on the spectral energy
density of scalar-induced, secondary GWs. 
We shall also utilize the PTA data---in particular, the NANOGrav 15-year 
data~\cite{NANOGrav:2023gor,NANOGrav:2023hde,Antoniadis:2023lym,EPTA:2023fyk,
Zic:2023gta, Reardon:2023gzh,Xu:2023wog}---to arrive at constraints on the 
reheating parameters.
Notably, we shall also arrive at the constraints while ensuring the 
corresponding $\fpbh(M)$ lie within the observational bounds that we 
mentioned above.


\section{Generation of scalar-induced secondary GWs and 
the PTA data}\label{sec:isgwb_pta}

As is well known, primordial GWs are the transverse and traceless components 
of the tensor perturbations in the metric describing the FLRW universe.
There can be different origins for GWs in the early universe. 
At the linear order in perturbation theory, the scalar and tensor perturbations
evolve independently.
However, the tensor perturbations at the second order can be sourced by the scalar 
perturbations at the first order.
As a result, whenever the amplitudes of the scalar perturbations are enhanced on small 
scales so that the PBHs are formed abundantly, they induce tensor perturbations of 
significant strengths at the second order, often leading to a peak in the spectral 
density of GWs (see, for instance, Refs.~\cite{Espinosa:2018eve,Kohri:2018awv}).
With the aim of explaining the PTA data, in this work, we focus on the generation
GWs that are associated with the formation of PBHs during the phase of reheating
described by a generic EoS. 
We shall consider an inflationary scalar power spectrum with a peak as given 
by Eq.~\eqref{eq:PR} and compare the resulting spectral energy density of induced 
GWs with the NANOGrav 15-year data. 
Since we are interested in understanding the compatibility of the GW background 
generated during reheating with the detected NANOGrav 15-year signal, we shall assume 
a considerably low reheating temperature of $\tre \simeq 50\, \mathrm{MeV}$.
Such an assumption ensures that the wave numbers with $k> \kre$ fall in the NANOGrav 
band of frequencies.
Further, it is known that, when $\wf>1/3$, a prolonged period of reheating can cause 
the spectral energy density of primary GWs (i.e. the GWs wave generated from the 
vacuum during inflation) to cross the so-called BBN bound. 
Later, we shall also utilize the bound to constrain the tensor-to-scalar ratio 
generated during inflation.


\subsection{Spectral density of secondary GWs over wave numbers that re-enter
during the phase of reheating}\label{sec:w}

As we mentioned earlier, when the primordial scalar power spectrum is boosted
on small scales to increase the number of PBHs formed, the enhanced scalar power
also induces secondary GWs of significant strengths.
In this section, we shall discuss the evaluation of the spectral energy density 
of the scalar-induced, secondary GWs generated during the phase of reheating and 
the epoch of radiation domination.
We shall closely follow the formalism developed in the earlier efforts in this context
(see, for instance, Refs.~\cite{Domenech:2019quo,Domenech:2020kqm,Domenech:2021ztg}).

Let $h_{ij}$ represent the tensor perturbations at the second order and 
let $h_{\vk}^\lambda$ denote the corresponding Fourier modes, where the 
superscript $\lambda=(+,\times)$ denotes the two types of polarization.
In the presence of a source, say, $S_{\vk}^\lambda$, the equation of motion 
governing the Fourier modes~$h_{\vk}^\lambda$ is given by
\begin{equation}
{h_{\vk}^\lambda}''+ 2\,a\,H\, {h_{\vk}^\lambda}'+k^2\, h_{\vk}^\lambda
=S_{\vk}^\lambda.\label{eq:eom-sgw}
\end{equation}
Recall that, in the absence of anisotropic stresses, at the first order, 
the scalar perturbations are described by the Bardeen potential~$\Phi$.
In a background characterized by the constant EoS parameter~$\wre$, the 
Fourier modes $\Phi_{\vk}$ of the Bardeen potential satisfy the following 
equation of motion:
\begin{equation}
\Phi_{\vk}''+3\,a\,H\,(1+\wf)\,\Phi_{\vk}' +\wf\, k^2\,\Phi_{\vk}=0.\label{eq:eom-bp}
\end{equation}
Since we are interested in the second-order tensor perturbations that are 
sourced by the first-order scalar perturbations, we can expect the source 
term~$S_{\vk}^\lambda$ to depend quadratically on the Fourier modes of the 
Bardeen potential~$\Phi_{\vk}$~\cite{Ananda:2006af,Baumann:2007zm,
Saito:2008jc,Saito:2009jt}.
At a time when the background is described by the constant EoS parameter~$\wre$,
the Bardeen potential is related to Fourier modes of the curvature perturbation~$\cR_{\vk}$ 
generated during inflation as follows:
\begin{equation}
\Phi_{\vk}=\f{3\,(1+\wre)}{5+3\,\wre}\,T(k\,\eta)\,\cR_{\vk},
\end{equation}
where $T(k\,\eta)$ is the so-called  transfer function that captures the 
time evolution of the Bardeen potential.  
One can show that the source term $S_{\vk}^\lambda$ can be expressed 
as~\cite{Baumann:2007zm,Saito:2008jc}
\begin{equation}
S_{\vk}^\lambda(\eta) 
= 4\l[\f{3\,(1+\wf)}{5+3\,\wf}\r]^2 \int \f{\d^3 {\bm p}}{(2\,\pi)^{3/2}}\,
\varepsilon^\lambda(\vk,\vp)\, f(p,q,\eta)\,\cR_{\vp}\,\cR_{\vq},
\end{equation}
where $\vq= (\vk- \vp)$ and, for convenience, we have introduced the quantity 
$\varepsilon^\lambda(\vk,\vp)=\varepsilon^\lambda_{ij}(\vk)\,p^i\,p^j$.
The quantity $f(p,q,\eta)$ involves combinations of the transfer function and 
its time derivative, and is given by
\begin{equation}
f(p,q,\eta) = 2\,T(p\,\eta)\, T(q\,\eta)
+\f{4}{3\,(1+\wre)}\, \l[T(p\,\eta)+\f{T'(p\eta)}{a\,H}\r]\,
\l[T(q\,\eta)+\f{T'(q\,\eta)}{a\,H}\r].\label{eq:f-tf}
\end{equation}
Also, it can be shown that 
\begin{equation}
\varepsilon^+(\vk,\vp)=\f{p^2}{\sqrt{2}}\,\sin^2\theta\, \cos (2\,\phi),\quad 
\varepsilon^\times(\vk,\vp)=\f{p^2}{\sqrt{2}}\, \sin^2\theta\, \sin(2\,\phi), 
\end{equation}
where $\theta$ is the angle between the wave vectors $\vk$ and $\vp$.  

Let $G_k(\eta,\te)$ denote the Green's function that satisfies the differential
equation
\begin{equation}
G_k''+\l(k^2-\f{a''}{a}\r)\,G_k=\delta^{(1)}(\eta-\te).\label{eq:eom-Gfn}
\end{equation}
In such a case, the Fourier modes $h_{\vk}^\lambda$ of the tensor perturbations 
at the second order, which are governed by Eq.~\eqref{eq:eom-sgw}, can be expressed 
in terms of the Green's function $G_k(\eta,\te)$ as follows:
\begin{equation} 
h_{\vk}^\lambda(\eta)
= 4 \int \f{\d^3 {\bm p}}{(2\,\pi)^{3/2}}\,\varepsilon^\lambda(\vk,\vp)\,
I(p,q,\eta)\,\cR_{\vp}\,\cR_{\vq},\label{eq:hk-lambda}
\end{equation}
where the kernel $I(p,q,\eta)$ is described by the integral
\begin{equation}
I(p,q,\eta) = \l[\f{3\,(1+\wre)}{5+3\,\wre}\r]^2\,
\int \f{\d\te\,a(\te)}{a(\eta)}\, G_k(\eta,\te)\, \,f(p,q,\te).
\end{equation}
As is well known, if $u_k^1(\eta)$ and $u_k^2(\eta)$ are the two solutions to a second
order, homogeneous differential equation, the Green's function associated with the 
inhomogeneous equation can be written as 
\begin{equation}
G_k(\eta,\te)=\f{u_k^1(\eta)\,u_k^2(\te)-u_k^1(\te)\,u_k^2(\eta)}{{u_k^1}'(\te)\,
u_k^2(\te) - u_k^1(\te)\, {u_k^2}'(\te)}.
\end{equation}
During the phase of reheating described by the constant EoS parameter $\wre$, 
the two homogeneous solutions to the differential equation~\eqref{eq:eom-Gfn} 
are given by $u_k^1(x) = \sqrt{x}\,J_{\alpha+1/2}(x)$ and $u_k^2(x)= \sqrt{x}\,
Y_{\alpha+1/2}(x)$, where $x=k\,\eta$, $\alpha=(1-3\,\wre)/(1+3\,\wre)$,
and $J_\nu(z)$ and $Y_\nu(z)$ are Bessel functions of the first and the second 
kind, respectively.
On using these solutions, the Green's function $G_k(\eta,\te)$ can be written
as
\begin{equation}
G_k(x,\tilde{x}) 
= -\f{\pi}{2\,k}\,\sqrt{{x}\,\tilde{x}} \, \l[J_{\alpha+1/2}(x)\,Y_{\alpha+1/2}(\tilde{x})
-J_{\alpha+1/2}(\tilde{x})\,Y_{\alpha+1/2}({x})\r].
\end{equation} 

We now require the transfer function $T(k\,\eta)$ describing the Bardeen potential
to construct the function~$f(p,q,\eta)$ [cf. Eq.~\eqref{eq:f-tf}].
During the phase of reheating, the general solution to Eq.~\eqref{eq:eom-bp} that 
governs the Bardeen potential can be expressed in terms of the Bessel functions
$J_\nu(z)$ and $Y_\nu(z)$  as follows:
\begin{equation} 
T(x) = x^{-(\alpha+3/2)}\,\l[C\, J_{\alpha+3/2}(\sqrt{\wre}\,x)
+D\, Y_{\alpha+3/2}(\sqrt{\wre}\,x)\r],
\end{equation} 
where $x=k\,\eta$, and the coefficients~$C$ and~$D$ are to be determined by matching 
the transfer function and its time derivative with the earlier epoch. 
Note that the limit $x\ll 1$ corresponds to early times.
In such a limit, since the second term in the above expression for $T(x)$ diverges, 
we shall set $D=0$, i.e. we shall assume that there are no decaying solutions.
Conventionally, the transfer function is chosen to be $T\to 1$ at large scales or,
equivalently, when $x\ll 1$.
Such a condition determines the value of the coefficient $C$ to be 
$C=(2/\sqrt{\wre})^{\alpha+3/2}\, \Gamma(\alpha+5/2)$ so that the
transfer function is given by
\begin{equation}
T({x}) = \l(\f{2}{\sqrt{\wre}\,x}\r)^{\alpha+3/2}\,
\Gamma\l(\alpha+\f{5}{2}\r)\,J_{\alpha+3/2}(\sqrt{\wre}\,x).\label{eq:tf}
\end{equation}
On using this expression for the transfer function and after some simplification,
we can express the function $f(p,q,\eta)$ as 
\begin{align}
f(u,v,{x}) 
&= \f{4^{\alpha+3/2}}{(2\,\alpha+3)(\alpha+2)}\,
\Gamma^2\l(\alpha+\f{5}{2}\r)\, (\wre\,u\,v\,x^2)^{-(\alpha+1/2)}\nn\\
&\quad\times\,\biggl[J_{\alpha+1/2}\l(\sqrt{\wre}\,u\,x\r)\, 
J_{\alpha+1/2}\l(\sqrt{\wf}\,v\,x\r)\nn\\
&\qquad+\l(\f{2+\alpha}{1+\alpha}\r) J_{\alpha+5/2}(\sqrt{\wre}\,u\,x)\,
J_{\alpha+5/2}(\sqrt{\wf}\,v\,x)\biggr],
\end{align}
where, for convenience, we have set $p=k\,v$ and $q=k\,u$. 

Given the energy density of GWs at any given time, say, $\rgw(\eta)$, the corresponding 
spectral energy density is defined as~$\rgw(k,\eta) =\d\rgw(\eta)/\d \ln k$.
The spectral energy density of the secondary GWs can be expressed as 
\begin{equation}
\rgw(k,\eta) =\f{\Mpl^2}{8}\,\l(\f{k}{a}\r)^2\,\overline{\mathcal{P}_h(k,\eta)},
\end{equation}
where $\cP_h(k,\eta)$ denotes the power spectrum of the GWs generated due to the second 
order scalar perturbations and is defined through the relation
\begin{equation}
\langle h_{\vk}^{\lambda}(\eta)\,h_{\vk'}^{\lambda'}(\eta)\rangle
=\f{2\,\pi^2}{k^3}\,\cP_h(k,\eta)\, \delta^{(3)}(\vk+\vk') \,\delta^{\lambda\,\lambda'}.
\end{equation}
We should clarify that the spectrum $\mathcal{P}_h(k,\eta)$ is to be evaluated at late 
times when the wave numbers of interest are well inside the Hubble radius and the overbar 
denotes the average over the oscillations that occur on small time scales.

Since the Fourier modes $h_{\vk}^{\lambda}$ depend quadratically 
on the Fourier modes~$\cR_{\vk}$ of the inflationary curvature 
perturbations [cf. Eq.~\eqref{eq:hk-lambda}], evidently, the power 
spectrum $\cP_h(k,\eta)$ will involve products of four~$\cR_{\vk}$.
On using Wick's theorem, which applies to Gaussian random variables, we 
can express the spectral density $\rgw(k,\eta)$ during reheating 
as~\cite{Domenech:2019quo,Domenech:2020kqm,Domenech:2021ztg}
\begin{align} 
\rgw(k,\eta) & = \f{\Mpl^2}{2} \l(\f{k}{a}\r)^2\,
\int_0^\infty\d v \int_{\vert 1-v\vert}^{1+v}\,\d u\, 
\l[\f{4\,v^2-\l(1+v^2-u^2\r)^2}{4\,u\,v}\r]^2\nn\\
&\quad\times\, \overline{I^2(u,v,x)}\, \cP_\cR(k\,u)\,\cP_\cR(k\,v),
\label{eq:ogw0}
\end{align}
\color{black}
where the kernel $I(u,v,x)$ can be expressed as
\begin{align}
I(u,v,x) &=  \f{4^{\alpha}\,(2+\alpha)\,\pi}{(3+2\,\alpha)\,
(\wre\,u\,v\,x)^{(\alpha+1/2)}}\;\Gamma^2\l(\alpha+\f{3}{2}\r)\nn\\
&\times\,\l[\cI_J(u,v,x)\,Y_{\alpha+1/2}({x})  
- \cI_Y(u,v,x)\,J_{\alpha+1/2}({x})\r]\label{eq:I1}
\end{align}
with the functions $\cI_J(u,v,x)$ and $\cI_Y(u,v,x)$ being described by
the integrals
\begin{subequations}
\begin{align} 
\cI_J(u,v,x) &= \int_0^{x} \d\tx\,\tx^{-\alpha+1/2}\, J_{\alpha+1/2}(\tx)\,
\biggl[J_{\alpha+1/2}(\sqrt{\wf}\,u\,\tx)\,J_{\alpha+1/2}(\sqrt{\wf}\,v\,\tx)\nn\\ 
&\quad+\l(\f{2+\alpha}{1+\alpha}\r)\,J_{\alpha+5/2}(\sqrt{\wre}\,u\,\tx)\,
J_{\alpha+5/2}(\sqrt{\wre}\,v\,\tx)\biggr],\\
\cI_Y(u,v,{x}) &= \int_0^{x} \d\tx\,\tx^{-\alpha+1/2}\, Y_{\alpha+1/2}(\tilde{x})\,
\biggl[J_{\alpha+1/2}(\sqrt{\wre}\,u\,\tx)\, J_{\alpha+1/2}(\sqrt{\wre}\,v\,\tx)\nn\\ 
&\quad +\l(\f{2+\alpha}{1+\alpha}\r)\,
J_{\alpha+5/2}(\sqrt{\wre}\,\tx)\,J_{\alpha+5/2}(\sqrt{\wre}\,\tx)\biggr].
\end{align}
\end{subequations}
While these integrals are difficult to calculate analytically for a finite~$x$, they 
can be evaluated in the limit $x\to \infty$.
It can be shown that, in such a limit, the integrals~$\cI_J(u,v,x)$ and 
$\cI_Y(u,v,x)$ can be expressed as
\begin{subequations}
\begin{align}
\lim_{x\to\infty}
\cI_{J}(u,v,x) &= \f{\pi}{2}\, \f{(\wre/2)^{\alpha-1/2}}{\pi\,\sqrt{\pi\, u\, v}}\, 
\Theta\l(u+v-\f{1}{\sqrt{\wre}}\r)\,Z^{\alpha}\,
\l[P^{-\alpha}_{\alpha}(y) +\l(\f{2+\alpha}{1+\alpha}\r)\,P^{-\alpha}_{\alpha+2}(y)\r],\\
\lim_{x\to\infty}
\cI_{Y}(u,v,x) &= -\f{(\wre/2)^{\alpha-1/2}}{\pi\,\sqrt{\pi\, u\,v}} 
\biggl\{\Theta\l( u+v-\f{1}{\sqrt{\wf}}\r)\, Z^{\alpha}\,
\l[Q^{-\alpha}_{\alpha}(y) +\l(\f{2+\alpha}{1+\alpha}\r)\, Q^{-\alpha}_{\alpha+2}(y)\r]\nn\\
&\quad- \Theta\l( \f{1}{\sqrt{\wf}}-u-v\r)\, \tilde{Z}^{\alpha}\, 
\l[\mathcal{Q}^{-\alpha}_{~\alpha}(\tilde y) 
+2\l(\f{2+\alpha}{1+\alpha}\r)\, \mathcal{Q}^{-\alpha}_{\alpha+2}(\tilde y)\r]\biggr\},
\end{align}
\end{subequations}
where
\begin{equation} 
y=\l(\f{u^2+v^2 - 1/\wf}{2\,u\,v}\r),\;\;
Z^2= 4\,u^2 v^2\,(1-y^2),\;\; \tilde{y}=-y,\;\; \tilde{Z}^2=-Z^2.
\end{equation}
In the above expressions, the functions $P^\mu_\nu(z)$ and $Q^\mu_\nu(z)$ are referred 
to as the Legendre polynomials on the cut, and $\mathcal{Q}^\mu_\nu(z)$ is the associated 
Legendre polynomial (we would refer the reader to Ref.~\cite{Domenech:2019quo} for 
further details).
In the late time limit of our interest, i.e. when $x\gg 1$, the Bessel functions in 
Eq.~\eqref{eq:I1} behave as $J_\nu(x)= \sqrt{2/(\pi\, x)}\,\cos\l[x-(2\,\nu+1)\,
\pi/4\r]$ and $Y_\nu(x)= \sqrt{2/(\pi\, x)}\,\sin\l[x-(2\,\nu+1)\,\pi/4\r]$. 
As a result, when we square $I(u,v,x)$ and average over the oscillations, the square 
of the sine and the cosine functions will be replaced by the average value of~$1/2$, 
while the cross term will vanish.
Therefore, we finally obtain that 
\begin{align} 
\lim_{x\gg 1} 
\overline{I^2(u,v,x)}
&\simeq \f{4^{\alpha}}{2\,\wre^2}\,
\l(\f{2+\alpha}{3+2\alpha}\r)^2\, \Gamma^4\l(\alpha+\f{3}{2}\r)\, 
(u\,v\,x)^{-2\,(\alpha+1)}\nn\\
&\quad\times \biggl\{\Theta\l(u+v-\f{1}{\sqrt{\wre}}\r)\, Z^{2\alpha}\,
\l[P^{-\alpha}_{\alpha}(y) 
+ \l(\f{2+\alpha}{1+\alpha}\r)\,P^{-\alpha}_{\alpha+2}(y)\r]^2\nn\\
&\qquad+\f{4}{\pi^2}\, \Theta\l(u+v-\f{1}{\sqrt{\wre}}\r)\,Z^{2\alpha}\,
\l[Q^{-\alpha}_{\alpha}(y) 
+\l(\f{2+\alpha}{1+\alpha}\r)\, Q^{-\alpha}_{\alpha+2}(y)\r]^2\nn\\
&\qquad+\f{4}{\pi^2}\,\Theta\l(\f{1}{\sqrt{\wre}}-u-v\r)\,
\tilde{Z}^{2\alpha}\, \l[\mathcal{Q}^{-\alpha}_{\alpha}(\tilde{y}) 
+2\l(\f{2+\alpha}{1+\alpha}\r)\, 
\mathcal{Q}^{-\alpha}_{\alpha+2}(\tilde{y})\r]\biggr\}\label{eq:I2}
\end{align}
and, with this quantity in hand, we can evaluate the integrals over $u$ and $v$
in Eq.~\eqref{eq:ogw0}, say, numerically, to eventually arrive at $\rgw(k,\eta)$.
From the above expression for the average of ${I^2(u,v,x)}$ and the factor of
$a^{-2}$ in Eq.~\eqref{eq:ogw0}, it should be clear that $\rgw(k,\eta)$ behaves 
as $a^{-4}$.
Such a behavior is over wave numbers that are well inside the Hubble radius at 
late times.
We should add that the $\rgw(k,\eta)$ we have obtained above applies to wave
numbers $\kre < k < \ke$, where, recall that, $\kre$ denotes the wave number 
that re-enters the Hubble radius at the end of the phase of reheating, while 
$\ke$ is the wave number that leaves the Hubble radius at the end of inflation
(or, equivalently, re-enters at the start of reheating).

Note that, since, in our analysis, we assume that the wave numbers near the 
peak in the scalar power spectrum re-enter during the phase of reheating (i.e.
they lie in the range $\kre < k < \ke$), the dominant contribution to the spectral 
density of secondary GWs can be expected to arise from the phase.
The phase of reheating will be followed by the epoch of radiation domination.
The secondary GWs generated over $\kre < k < \ke$ during reheating will freely
propagate during radiation domination.
In the following section, we shall discuss the generation of secondary GWs 
during the epoch of radiation domination.


\subsection{Spectral density of secondary GWs over wave numbers that re-enter
during radiation domination}

Let us now turn to wave numbers $k<\kre$ that re-enter the Hubble radius 
during the epoch of radiation domination.
Given that the radiation-dominated epoch was preceded by the phase of reheating,
the general solution for the transfer function $T(x)$ for the scalar perturbations
at the first order can be written as~\cite{Inomata:2019ivs}
\begin{align}
T(x) & =\f{3\, \sqrt{3}}{x-\alpha\,\xre/(1+\alpha)}\,
\biggl\{A(\xre)\, j_1\l[\f{1}{\sqrt{3}}\,\l(x-\f{\alpha}{1+\alpha}\,\xre\r)\r]\nn\\
&\quad +B(\xre)\,y_1\l[\f{1}{\sqrt{3}}\,\l(x-\f{\alpha}{1+\alpha}\,\xre\r)\r]\biggr\},
\label{eq:phik-rd}
\end{align}
where $j_1(z)$ and $y_1(z)$ are the spherical Bessel functions of the first and 
second kind, and $\xre= k\,\ere$. 
Note that the above transfer function accounts for the redefinition of the scale factor
due to the presence of a reheating phase, in contrast to the case of instantaneous 
reheating [see $a(\eta)$ in Eq.~\eqref{eq:rd}].
This aspect needs to be taken into account carefully while deriving the kernel 
$I(u,v,x)$ for the epoch of radiation domination. 
Needless to say, the effect of this redefinition becomes unimportant for wave 
numbers such that $k \ll \kre$, i.e. modes which re-enter during the time $\eta 
\gg \ere$.

There is an important point that we need to clarify at this stage of our discussion.
Consider a sharply peaked primordial scalar power spectrum with its maximum at~$\kp$.
If the wave number $\kp$ re-enters during the phase of reheating and, if $\kre\simeq 
\kp$, then, due to the nature of the convolution integral describing the spectral 
density of secondary GWs, the GWs generated over $k \lesssim \kre$ can also be affected 
by the phase of reheating (in this regard, see Refs.~\cite{Domenech:2020kqm,Domenech:2021ztg}). 
Such an effect can also arise in the scenario that we consider in this work. 
It is precisely to avoid such an effect that we have assumed a rather low reheating
temperature, which leads to a low value for $\kre$ thereby ensuring that $\kre \ll \kp$. 
In such a situation, the effects of the modification to the kernel $I(u,v,x)$ during
radiation domination due to the redefined scale factor do not play a significant role 
in estimation of the spectral density of secondary GWs.

The more general form of the kernel during radiation domination---which accounts for
the prior phase of reheating---can be derived with $\xre$ redefined as $\xre/2 \to 
[\alpha/(1+\alpha)]\,\xre$ (in this regard, see Ref.~\cite{Bhaumik:2020dor}, App.~A).
Moreover, the constants $A(\xre)$ and $B(\xre)$ in the transfer function~\eqref{eq:phik-rd} 
above can be determined by matching the Bardeen potential and its time derivative 
at~$\ere$. 
Clearly, the resulting kernel will reduce to the standard kernel in the epoch 
of radiation domination for wave numbers $k \ll \kre$ that re-enter much later 
than~$\ere$.
In the late time limit, i.e. as $x \to \infty$, we can write the kernel during
radiation domination as~\cite{Espinosa:2018eve,Kohri:2018awv}
\begin{align}
\lim_{x\to\infty}\overline{I^2(v,u,x)}
& =  \f{1}{2}\,\l[\f{3\, (u^2+v^2-3)}{4\,u^3\,v^3\,x}\r]^2\,
\biggl\{\l[4\,u\,v +(u^2+v^2-3)\,\mathrm{log}\l|\f{3-(u+v)^2}{3-(u-v)^2}\r|\r]^2\nn\\
&\quad + \Theta\l(u+v-\sqrt{3}\r)\,
\pi^2\, (u^2+v^2-3)^2\biggr\}.\label{ird}
\end{align}
In our scenario, which involves a sharply ascending scalar power spectrum 
with a peak that re-enters the Hubble radius well before the epoch of
radiation domination, we can make use of this kernel in Eq.~\eqref{eq:ogw0} 
without losing any accuracy in the estimate of the spectral density of 
secondary GWs over wave numbers such that $k \ll \kre < \kp$.

The dimensionless spectral energy density parameter $\ogw(k,\eta)$ associated 
with $\rgw(k,\eta)$ is defined with respect to the critical density at 
the time, viz. $\rho_\mathrm{cr}(\eta)$, as 
\begin{equation}
\ogw(k,\eta) =\f{\rgw(k,\eta)}{\rho_{\mathrm{cr}}(\eta)}
= \f{\rgw(k,\eta)}{3\,\Mpl^2\, H^2}.\label{Omega_1}
\end{equation}
The present-day, dimensionless spectral energy density of secondary GWs can be
expressed in terms of the quantity $\ogw(k,\eta)$ as follows:
\begin{equation}
\ogw(k)\, h^2 =  c_g\,\oR\, h^2\,\ogw(k,\eta) 
=\f{c_g\,\oR\, h^2}{24}\, \l(\f{k}{a\,H}\r)^2\,\overline{\cP_h(k,\eta)},
\end{equation}
where $c_g=\l(g_k/g_0\r) \l(g_{\mathrm{s,}0}/g_{\mathrm{s},k}\r)^{4/3}$,
with $(g_k,g_{\mathrm{s},k})$ and $(g_0,g_{\mathrm{s},0})$ denoting the 
effective number of relativistic degrees of freedom associated with the 
energy density of radiation and entropy when the wave number $k$ re-enters 
the Hubble radius and today, respectively. 

\begin{figure}[!t]
\centering
\includegraphics[width=0.9\linewidth,height=9cm]{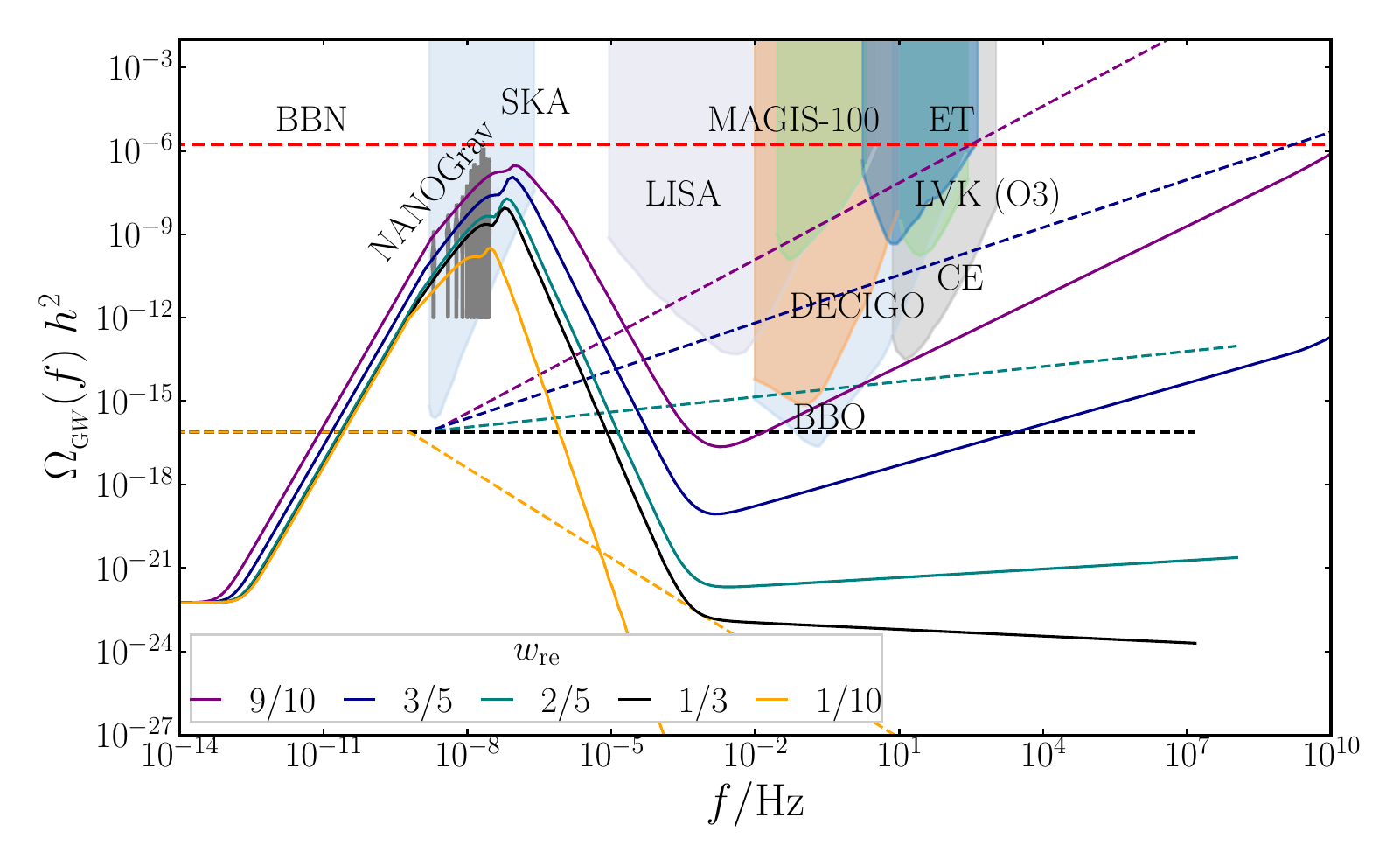}
\caption{The dimensionless spectral energy density~$\ogw$ of both the primary 
(as dashed curves) and secondary (as solid  curves) GWs arising in the scenario
of interest are plotted as a function of the frequency~$f$.
We have plotted the spectral energy densities up to the
frequency corresponding to the wavenumber~$\ke$ which leaves the Hubble radius 
at the end of inflation.
In plotting the spectral density of secondary GWs, we have chosen the parameters 
of the primordial scalar power spectrum [cf. Eq.~\eqref{eq:PR}] to be $\log_{10}A_0
=-1.5$, $\log_{10}(\kp/\mathrm{Mpc}^{-1})=7.5$ and $n_0=-2$.
We have set the reheating temperature to be~$\tre=50\, \mathrm{MeV}$ and have 
plotted the results for the following values of the EoS parameter during 
reheating: $\wre=(1/10, 1/3, 2/5, 3/5, 9/10)$.
We should mention that the spectral energy densities of the primary GWs are plotted 
corresponding to the maximum value of the tensor-to-scalar ratio allowed by the CMB
data, viz. $r=0.036$~\cite{BICEP:2021xfz}, which sets the upper bound on the energy 
scale of inflation (in this context, see our discussion in Sec.~\ref{sec:pgw}).
In the figure, we have also indicated the so-called BBN bound (as the dashed red
line), which limits the total energy density of relativistic fields during the 
epoch of radiation domination.}\label{fig:ogw1}
\end{figure}

In Fig.~\ref{fig:ogw1}, we have plotted the dimensionless spectral 
density~$\ogw$ of the primary (in this context, see Sec.~\ref{sec:pgw}) 
and secondary GWs as a function of the frequency~$f$.
In the figure, we have also included the projected sensitivity curves of
different current and forthcoming GW observatories such as PTA, Square
Kilometre Array (SKA)~\cite{Moore:2014lga}, 
LISA~\cite{Bartolo:2016ami}, 
DECIGO~\cite{Kawamura_2011, Kawamura:2019jqt}, 
BBO~\cite{Crowder:2005nr,Corbin:2005ny,Baker:2019pnp}, 
MAGIS-100~\cite{Espinosa:2018eve, Coleman:2018ozp}, 
advanced~LIGO + Virgo (LVK)~\cite{LIGOScientific:2016jlg,LIGOScientific:2019vic},
ET~\cite{Sathyaprakash:2012jk}, 
and CE~\cite{LIGOScientific:2016wof}. 
We should point out that the sensitivity curves have been arrived at by 
assuming a power law for the spectral density of GWs and integrating the 
noise data over a range of frequencies and observation time.
The sensitivity curves indicate in a simple way whether the spectral 
densities we have calculated will be detected by the GW observatories.
Note that, in plotting the dimensionless spectral energy densities of the primary 
and secondary GWs, we have fixed the reheating temperature to be $\tre \simeq 50\,
\mathrm{MeV}$.
Moreover, in the case of the secondary GWs, we have assumed that the scalar 
power spectrum has a broken power law form, as given in Eq.~\eqref{eq:PR}.
Further, we have chosen that the peak in the scalar power spectrum to be 
at $\kp=10^{7.5}\, \mathrm{Mpc}^{-1}$, and we have assumed that~$n_0=-2$.
The reason for choosing these values of $\tre$ and $\kp$ is to effectively
constrain~$\wf$ with the aid of the PTA data.
Recall that, we have assumed a sudden transition from the phase of reheating
to the era of radiation domination.
We have chosen the above-mentioned parameters such that the wave number $\kre$
occurs before the range of frequencies that the PTAs are sensitive to.
In the figure, we have plotted the quantity~$\ogw(f)$ for the following five 
values of the parameter that describes the EoS during reheating: $\wf=(1/10, 
1/3, 2/5, 3/5, 9/10)$.
It is clear from the figure that, as $\wf$ increases, the height of the peak 
in the secondary~$\ogw(f)$ also increases and the bump-like feature around
the peak is flattened. 
It is also evident from the figure that, for all values of $\wf$, prior to 
the peak, the spectral density of the secondary GWs behaves as expected for
wave numbers that re-enter the Hubble radius during the epoch of radiation
domination.
Beyond the peak, the spectral density of the secondary GWs decreases in 
different manner for different values of~$\wf$. 
As $\wre$ is increased from zero, $\ogw(f)$ falls less and less and it becomes 
nearly constant for $\wre=1/3$, and starts increasing again for~$\wre>1/3$. 
This happens due to the relative redshift between the background and GWs.
The second rise in the spectral density of secondary GWs can also be detected 
by observatories such as LISA, DECIGO, and BBO. 
Hence, the presence of a double peak can indicate an EoS with~$\wf>1/3$.

There are a couple of related points regarding Fig.~\ref{fig:ogw1} that require
further emphasis.
In the figure, we have plotted the spectral energy densities of the primary and 
secondary GWs up to the frequency corresponding to the wavenumber~$\ke$, where,
recall that, $\ke$ is the wavenumber that leaves the Hubble radius at the end 
of inflation.
Note that the Fourier modes with wavenumbers larger than $\ke$ always stay within 
the Hubble radius and hence they do not contribute to the energy density of GWs 
(for a discussion in this regard, see, for example, Refs.~\cite{Giovannini:2008tm,
Pi:2024kpw}). 
It is clear from the figure that, for $\wre=9/10$, the spectral energy density 
of primary GWs crosses the sensitivity curve of LVK as well as the BBN bound 
from~$\dneff$.
Whereas, for $\wre=3/5$, the spectral energy density of the primary GWs crosses 
only the BBN bound. 
Based on the fact that LVK did not observe a SGWB~\cite{KAGRA:2021kbb}, we can 
exclude the spectral energy densities passing through it.
Therefore, for the tensor-to-scalar ratio of $r=0.036$, we can rule out the possibility
of a prolonged reheating phase with either~$\wre=9/10$ or~$\wre=3/5$.
But, a smaller value of $r$ (or, equivalently, a lower inflationary energy scale) 
would still remain compatible with such reheating histories. 
These suggest that, for a given energy scale of inflation, BBN and LVK can also 
be used to constrain the parameters that characterize the epoch of reheating.
In fact, blue-tilted inflationary spectra can be constrained by LVK very effectively. 
However, we should point out that, if we consider a flat or red-tilted inflationary
tensor power spectrum, the bound from $\dneff$ provides a stronger constraint on 
primary GWs than LVK.


\subsection{Comparison with the  NANOGrav 15-year data}

In this section, we scan our parameter space to find the region of the 
space that optimally fits the PTA data. 
We shall further examine whether the optimal values for the parameters we 
obtain are consistent with the available constraints on~$\fpbh(M)$.

\begin{figure}[!t]
\includegraphics[width=0.475\linewidth]{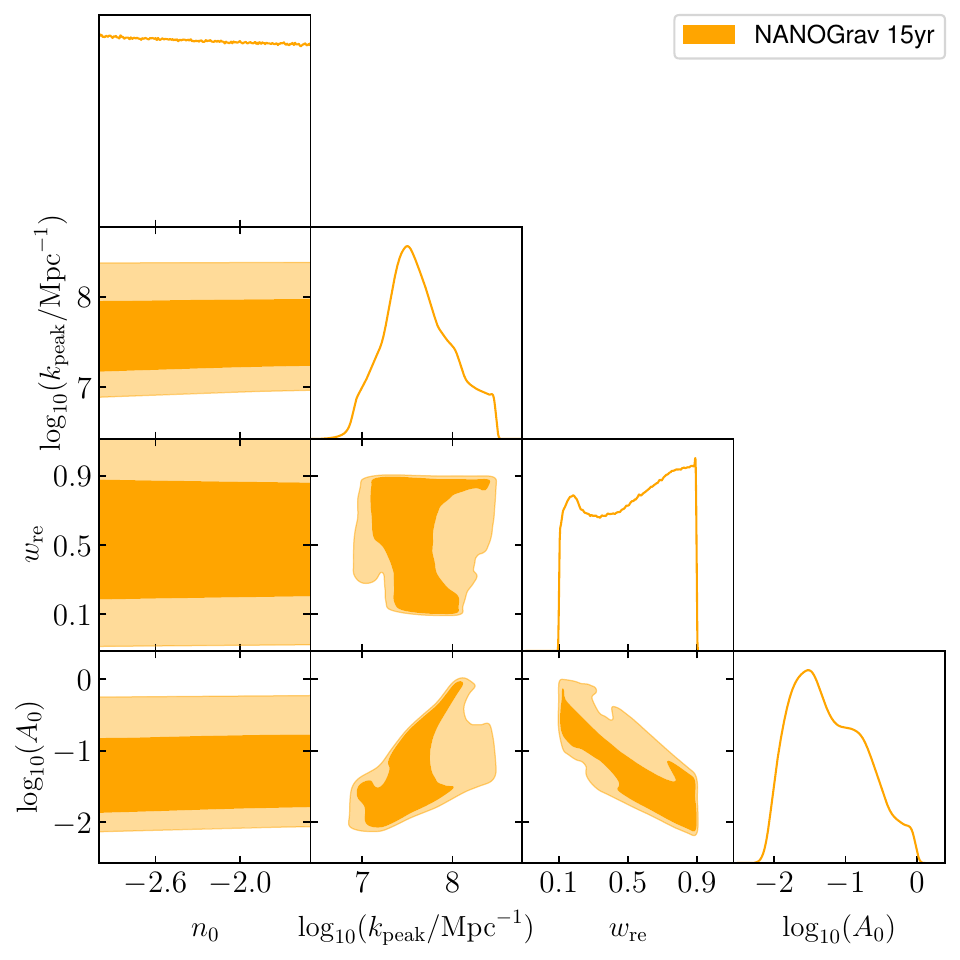} 
\includegraphics[width=0.475\linewidth]{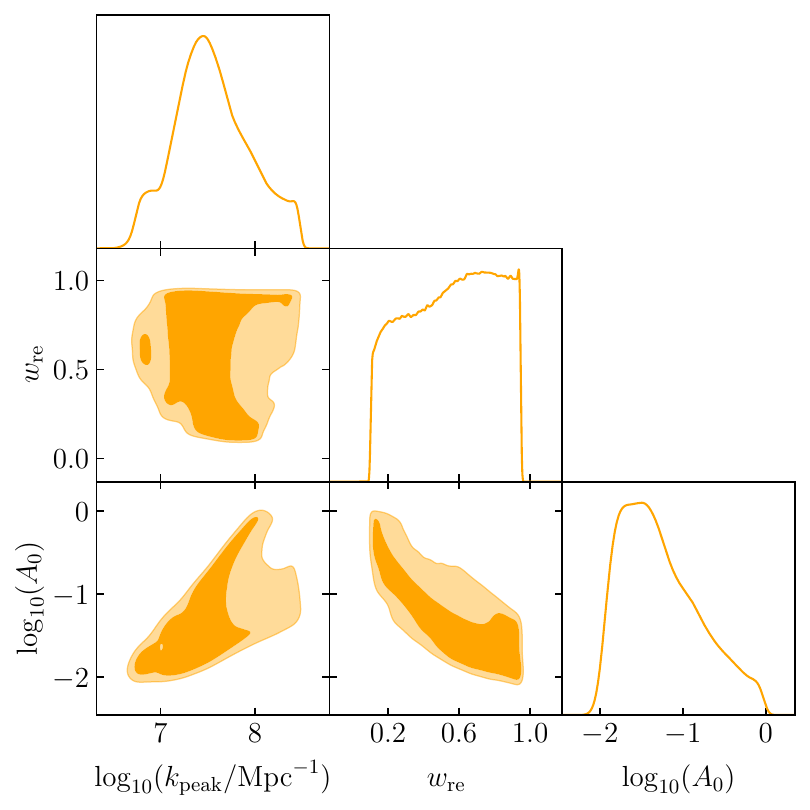} 
\caption{We have plotted the marginalized posterior distributions of the 
parameters that have been arrived at upon comparing the models {\tt R4pF} 
(on the left) and {\tt R3pF} (on the right) with the NANOGrav 15-year 
data.
Note that, in both the models, while the data constrains the parameters 
$A_0$ and $\kp$ fairly well (as is suggested by the peaks in their posterior 
distributions), the constraints on the parameters $\wre$ and $n_0$ (in
the case of the model~{\tt R4pF}) are rather weak. 
In fact, in the case of~$n_0$, the posterior distribution
is flat over the range of priors that we have worked with
[cf. Tab.~\ref{tab:upop}].}\label{fig:4p_3p}
\end{figure}

We use the NANOGrav 15-year data~\cite{NANOGrav:2023gor,NG15yrdata} and, 
to arrive at the best-fit parameters, we carry out a Markov-Chain Monte 
Carlo (MCMC) analysis with the help of ${\tt PTArcade}$~\cite{Mitridate:2023oar} 
in the~${\tt ceffyl}$ mode. 
Our initial search consists of four parameters, viz. the three parameters $(A_0,\kp, 
n_0)$ that describe the inflationary scalar power spectrum [cf. Eq. \eqref{eq:PR}] 
and the parameter~$\wf$ that describes the EoS during reheating. 
As we discussed earlier, our method to calculate the spectral density of 
secondary GWs lose the desired precision around wave numbers that re-enter
the Hubble radius close to the time of transition from the phase of reheating
to the epoch of radiation domination.
To circumvent this limitation, we choose the reheating temperature to be $\tre =50\, 
\mathrm{MeV}$, so that the frequency corresponding to $\kre$ always remains much 
smaller than the frequency range probed by NANOGrav. 
First, we run {\tt PTArcade} to sample over the four parameters $(A_0,\kp,n_0,\wf)$ 
and, for convenience, we refer to this model as~{\tt R4pF}. 
In Tab.~\ref{tab:upop}, we have listed the priors we work with and the mean 
values we have obtained from the run.
\begin{table}[!t]
\begin{center}
\begin{tblr}{|c|c|c|c|c|c|}
\hline  
\bf{Model} & \bf{Parameter} & \bf{Prior} & \SetCell[c=3]{c}{\bf{Mean value}} & &\\
\hline 
\SetCell[r=4]{c}{\tt{R4pF}}
& $\log_{10}\l(\f{\kp}{\mathrm{Mpc}^{-1}}\r)$ & $[6, 9]$ &
\SetCell[c=3]{c}$7.62^{+0.35}_{-0.41}$ & &\\
\hline
& $\log_{10}(A_{0})$ & $[-3, 0]$ &\SetCell[c=3]{c}$-1.23^{+0.38}_{-0.66}$ & &\\
\hline
& $\wre$ & $[0.1, 0.9]$ &\SetCell[c=3]{c}$0.52\pm 0.23$ & & \\
\hline
& $n_0$ &  $[-3.0, -1.5]$ &\SetCell[c=3]{c}$-2.26\pm 0.43$ & & \\
\hline
\SetCell[r=3]{c}{\tt{R3pF}} & $\log_{10}\l(\frac{\kp}{\mathrm{Mpc}^{-1}}\r)$ 
& $[6, 9]$ &\SetCell[c=3]{c}$7.54^{+0.36}_{-0.44}$ & &\\
\hline
& $\log_{10}(A_{0})$ & $[-3, 0]$ &\SetCell[c=3]{c}$-1.26^{+0.26}_{-0.64}$ & & \\
\hline
& $\wre$ & $[0.1, 0.9]$ &\SetCell[c=3]{c}$0.55^{+0.39}_{-0.14}$ & &\\
\hline  
& & & {\boldmath $0.5\, \dcan$} &  {\boldmath $\dcan$} & {\boldmath $1.5\,\dcan$}\\
\hline
\SetCell[r=3]{c}{\tt{R3pB}} & $\log_{10}\l(\f{M}{M_{\odot}}\r)$ & $[-6, 3.5]$ 
& $-0.12^{+0.28}_{-0.15}$  &$-1.18^{+0.35}_{-0.39}$&$-1.85^{+0.49}_{-0.30}$\\
\hline
& $\log_{10}(\fpbh)$ & $[-20, 0]$ & $-0.67^{+0.68}_{-0.16}$&$-6.6^{+6.5}_{-1.9}$&$-10.2^{+8.2}_{-9.6}$\\
\hline
& $\wre$ & $[0.1, 0.9]$ &$0.78^{+0.11}_{-0.030}$&$0.66^{+0.23}_{-0.19}$&$0.55\pm 0.17$\\
\hline
\SetCell[r=2]{c}{\tt{R2pB}} & $\log_{10}\l(\f{M}{M_{\odot}}\r)$ & $[-6, 3.5]$ 
& $-0.24^{+0.38}_{-0.45}$&$-1.60^{+0.16}_{-0.14}$&$-2.45^{+0.20}_{-0.13}$\\
\hline
& $\wre$ & $[0.1, 0.9]$ &$0.77^{+0.13}_{-0.038}$&$0.59\pm 0.16$&$0.464^{+0.095}_{-0.25}$\\
\hline
\end{tblr}
\caption{We have listed the parameters in the four models we consider, viz.
{\tt R4pF}, {\tt R3pF}, {\tt R3pB} and {\tt R2pB}, the priors we have worked 
with, and the mean values we have obtained upon comparison of the models with 
the NANOGrav 15-year data.
We should mention that, in our analysis, we have assumed the priors to be uniform.}
\label{tab:upop}
\end{center}
\end{table}
And, in Fig.~\ref{fig:4p_3p}, we have illustrated the resulting marginalized 
posterior distributions (known, popularly, as the triangular plots). 
It is evident from the posterior distributions that the PTA data does not 
constrain the parameter~$n_0$.
Such a result should not be surprising since the NANOGrav 15-year data only
favors an ascending spectral density of GWs~$\ogw(f)$ over the range of 
frequencies it is sensitive to.
In fact, a nearly scale-invariant spectrum is ruled out beyond $3$-$\sigma$ 
by the data.
This, in turn, implies that the data constrains the rising part of the 
inflationary scalar power spectrum prior to the peak. 
We should add that, in the model {\tt R4pF}, we do not compute the mass fraction 
$\fpbh(M)$ or do not take into account the constraints on the abundance of PBHs.
It would be interesting to explore the effects of changing $n_0$ on the PBH mass 
fraction and the resulting constraints from the NANOGrav 15-year data, particularly
when one takes into account the dependence of~$\dc$ on the shape of the inflationary
scalar power spectrum~\cite{Germani:2018jgr,Musco:2020jjb}. 
However, for the present work, we shall assume the effects of changing~$n_0$ to be 
negligible and, as the next step, we fix the value of $n_0$ to consider a model that
involves only the three parameters $(A_0, \kp,\wf$).
We refer to this three parameter model as $\tt{R3pF}$. 
As before, in Tab.~\ref{tab:upop}, we have listed the priors and best-fit values 
arrived at upon comparison with the NANOGrav 15-year data. 
Also, in Fig.~\ref{fig:4p_3p}, we have illustrated the marginalized posterior 
distributions that we have obtained.  

\begin{figure}[t!]
\includegraphics[width=0.475\linewidth]{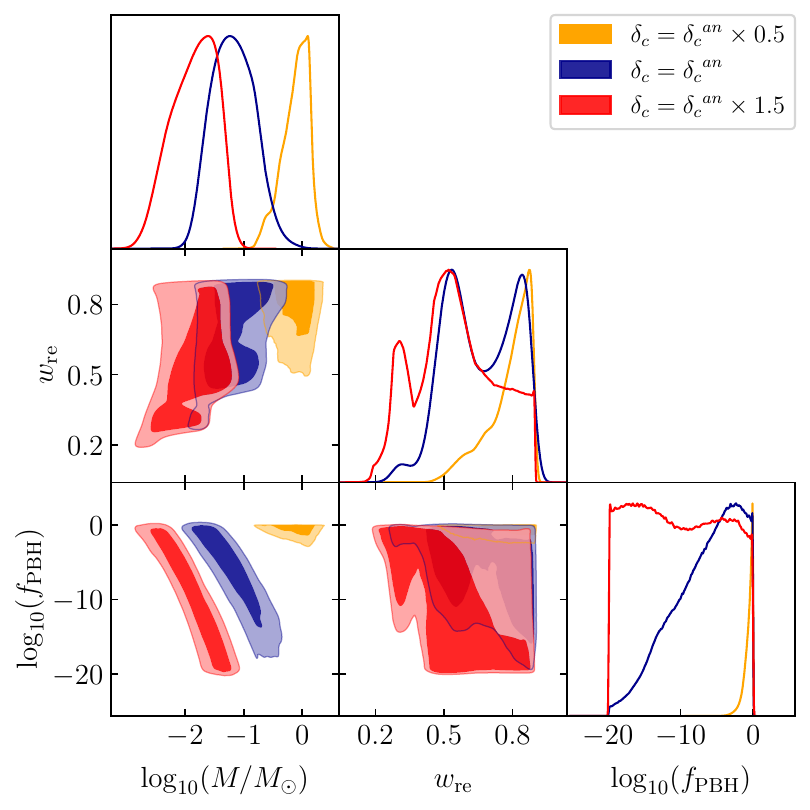} 
\includegraphics[width=0.475\linewidth]{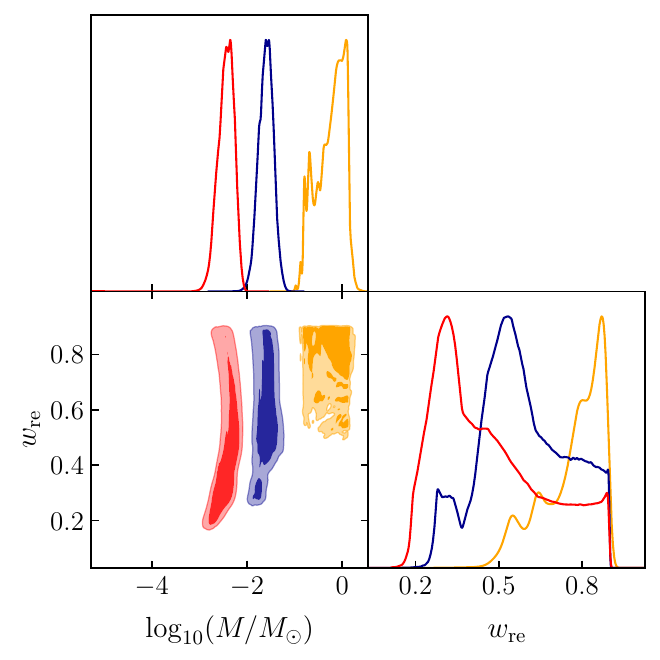} 
\caption{As in the previous figure, we have plotted the marginalized posterior
distributions of the parameters that have been arrived at upon comparing the 
models {\tt R3pB} (on the left) and {\tt R2pB} (on the right) with the NANOGrav 
15-year data.
A notable feature of these results is that, for lower values of~$\dc$, the data 
prefers higher values of~$\mbh$ and~$\wre$.}
\label{fig:3p_2p}
\end{figure}

Until this point, we have only compared the two models~{\tt R4pF} and~{\tt R3pF}
with the NANOGrav 15-year data.
We also need to make sure that the best-fit parameters that describe the scalar 
power spectrum (in particular, the amplitude~$A_0$) does not lead to the overproduction
of PBHs, an issue that has been encountered recently in similar contexts in the
literature (in this regard, see, for instance, Refs.~\cite{Inomata:2023zup,
Harigaya:2023pmw,Liu:2023pau,Zhao:2023joc}). 
To ensure that there is no excessive production of PBHs, we carry out fresh runs 
wherein we directly sample the parameters that describe the PBHs, viz.~$M$ 
and~$\fpbh$, along with the reheating parameter~$\wf$. 
In other words, for a given set of parameters~$(A_0,\kp,\wf)$, we calculate the 
quantity $\fpbh(M)$ and ensure that it is always less than unity.
As we discussed in Sec.~\ref{secPBH}, it is the uncertainties related to the collapse 
of PBHs that play an important role in determining the value of the critical density 
contrast~$\dc$. 
In order to understand the effects of the uncertainties on our results, apart
from the value of the critical density $\dc^{\mathrm{an}}$ given by the analytical 
expression~\eqref{eq:cd}, we also scan the parameter space for a smaller ($\dc =
0.5\,\dcan$) and a higher ($\delta_c=1.5\,\dcan$) value of~$\dc$. 
We refer to the model as~{\tt R3pB}. 
We indeed find that the change in the value of $\dc$ has a strong effect on the
optimal or best-fit values.
Such a conclusion should be evident from Tab.~\ref{tab:upop} wherein we have
listed the optimal values, and the posterior distributions we have plotted in 
Fig.~\ref{fig:3p_2p}.
The effect is particularly clear from the posterior distribution for $\fpbh$ plotted
in Fig.~\ref{fig:3p_2p}.
Note that, for $\dc=0.5\,\dcan$, the PBHs are overproduced for almost the entire 
range of the reheating parameter~$\wre$, barring values close to unity.
This is understandable as a larger~$\wre$ implies a larger~$\dcan$.
In complete contrast, for $\dc=1.5\,\dcan$, the entire range of $\fpbh$ we consider 
is allowed with equal probability for almost all values of~$\wf$.
If we choose the intermediate value of $\delta_c=\dcan$, we find that the condition 
that PBHs are not overproduced does not significantly constrain~$\wf$.
However, we do find that the data favors $\wre \gtrsim 0.5$.
Moreover, the posterior distribution for $\fpbh$ shows a monotonic rise with the 
increasing $\fpbh$.

\begin{figure}[!t]
\centering
\includegraphics[scale=.5]{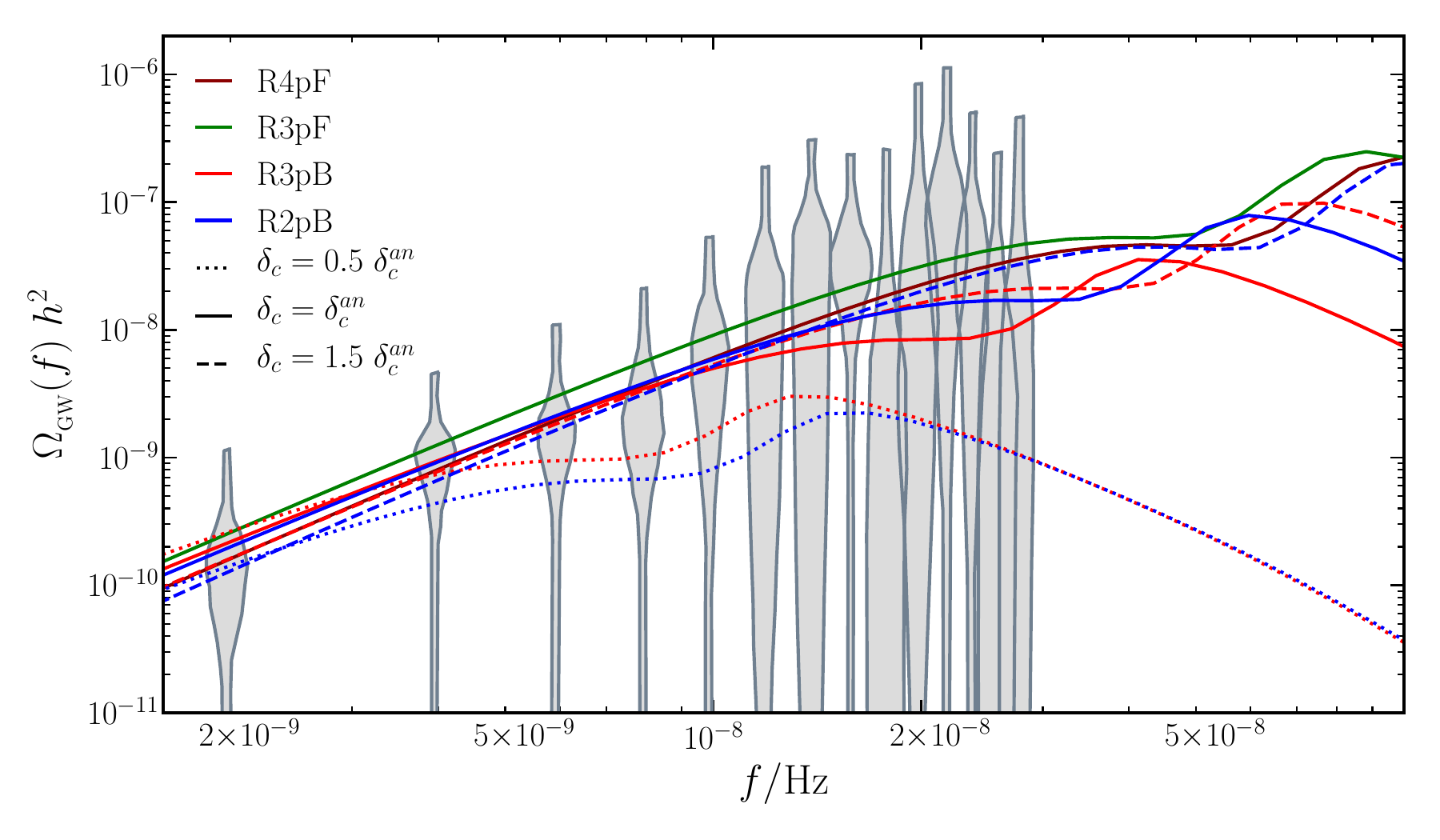}    
\caption{The spectral energy density of secondary GWs  $\ogw(f)$ that correspond 
to the best-fit values of the parameters in the four models, viz. {\tt R4pF}, 
{\tt R3pF}, {\tt R3pB} and {\tt R2pB}, are plotted.
In the figure, we have also included the probability densities from the Kernel 
Density Estimators (KDEs) of the NANOGrav 15-year data (as gray 
violins)~\cite{the_nanograv_collaboration_2023_8060824}.
It is visually evident that, when $\dc=0.5\,\dcan$, the spectral energy densities 
corresponding to the latter two models do not fit the data well.}
\label{fig:ogwbest1}
\end{figure}

The condition that $\fpbh < 1$ is the theoretical bound that exists on all 
mass scales.
But, there are also mass-dependent constraints that need to be taken into
account in the range of optimal PBH masses that we obtain. 
As illustrated in Fig.~\ref{fig:ogwbest1}, in the range of PBH masses of our interest, 
the constraints through micro-lensing from HSC OGLE~\cite{Niikura:2017zjd,Croon:2020ouk, 
Niikura:2019kqi} and the constraints from the Planck-2018 CMB data~\cite{Planck:2018jri} 
play the most important role. 
Therefore, in our final run, we fix~$\fpbh(M)$ to the maximum possible value 
allowed by the constraints from the different micro-lensing observations (as 
shown in Fig.~\ref{fig:ogwbest1}), and scan only over the parameters~$M$ 
and~$\wf$. 
We refer to this model as {\tt R2pB}.
Further, we perform a Bayesian analysis of the two-parameter model. 
As in the case of the model {\tt R3pB}, we allow $\dc$ to take three different 
values and compare them with the PTA data.
And, as in the earlier model, we find clear distinction in the optimal parameter
range in these different cases.  

To carry out a Bayesian analysis of the two-parameter model~{\tt R2pB} against
the NANOGrav 15-year data, we make use of {\tt PTArcade}~\cite{Mitridate:2023oar} 
in the {\tt enterprise}~\cite{enterprise} mode without the Hellings and 
Downs~\cite{1983ApJ...265L..39H} correction. 
Let $\mathcal{D}$ denote the PTA data and let $\theta$ denote the parameters
describing our model.
As is well known, in such a case, the posterior 
distribution~$\mathcal{P}(\theta|\mathcal{D})$ is given by
\begin{equation}
\mathcal{P}(\theta|\mathcal{D}) 
= \f{\mathcal{P}(\mathcal{D}|\theta)\,\mathcal{P}(\theta)}{\mathcal{P}(\mathcal{D})},
\end{equation}
where $\mathcal{P}(\theta)$ is the distribution of the priors on the parameters
involved, and $\mathcal{P}(\mathcal{D})$ denotes the marginalized likelihood.
We obtain the marginalized likelihood in support of model~$Y$ and utilize it to 
evaluate the Bayesian factor against a reference model~$X$ through the 
relation
\begin{equation}
\mathrm{BF}_{Y,X} \equiv \f{\mathcal{P}(\mathcal{D}|Y)}{\mathcal{P}(\mathcal{D}|X)}.
\end{equation}
For the reference model X, we choose the source of GWs to be the merging of 
SMBHBs.
In~{\tt PTArcade} and~{\tt enterprise}~\cite{enterprise}, we choose the
{\tt smbhb=True} and {\tt bhb\_th\_prior=True} options, which assumes
two-dimensional Gaussian priors for the parameters 
$A_{\scriptscriptstyle\textrm{BHB}}$ and 
$\gamma_{\scriptscriptstyle\textrm{BHB}}$
of the SMBHBs~\cite{NANOGrav:2023hvm,Mitridate:2023oar}.
We have listed the Bayesian factors for the model~{\tt R2pB} corresponding to the 
three different choices of $\dc$ in Tab.~\ref{tab:BF}.
\begin{table}
\begin{center}
\begin{tblr}{|c|c|c|c|c|}
\hline
\SetCell[r=2]{c}{{{\textbf{Model~X}}}}
& \SetCell[r=2]{c}{{{\textbf{Model~Y}}}} &  
\SetCell[c=3]{c}{{{${\mathrm{BF}_{Y,X}}$}}}\\
&& \hline
$\dc=0.5\,\dcan$ & $\dc=\dcan$ & $\dc=1.5\,\dcan$\\
\hline
SMBHBs & {\tt R2pB}  & $1.7\pm .06$ & $260.04\pm 19.21$ 
& $350.61\pm 27.36$\\
\hline
\end{tblr}
\caption{We have listed the Bayesian factors $\mathrm{BF}_{Y,X}$ that we have 
obtained for the model~{\tt R2pB} which invokes primordial physics as the 
source of the SGWB observed by the NANOGrav 15-year 
data, when compared to the astrophysical scenario of SMBHBs.  
Bayesian factors $\mathrm{BF}_{Y,X}$ that far exceed unity indicate 
strong evidence for the model~$Y$ with respect to the model~$X$. 
Clearly, when $\dc=\dcan$ and $\dc=1.5\,\dcan$, the NANOGrav 15-year
data strongly favors the model~{\tt R2pB} when compared to the model
of SMBHBs.}\label{tab:BF}
\end{center}
\end{table}
It is clear from the table that, when $\dc=0.5\, \dcan$, the model~{\tt R2pB} 
performs only slightly better than the model of SMBHBs.
However, when $\dc=\dcan$ and $\dc=1.5\,\dcan$, our comparison with the 
NANOGrav's 15-year data finds strong Bayesian evidence in favor of the 
scenario wherein PBHs are formed during reheating, resulting in the 
generation of secondary GWs rather than the SMBHB model.

\begin{figure}[!t]
\centering
\includegraphics[width=0.495\linewidth]{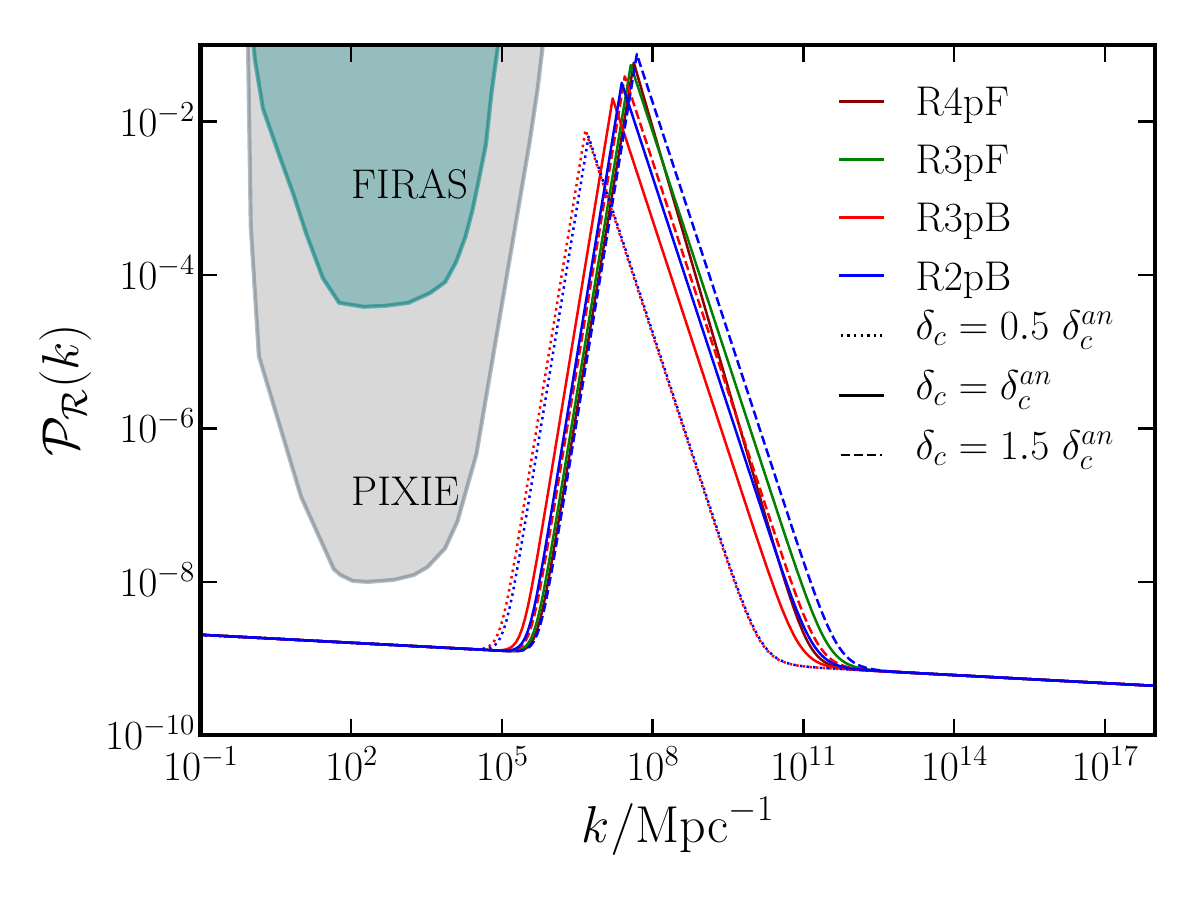}
\includegraphics[width=0.495\linewidth]{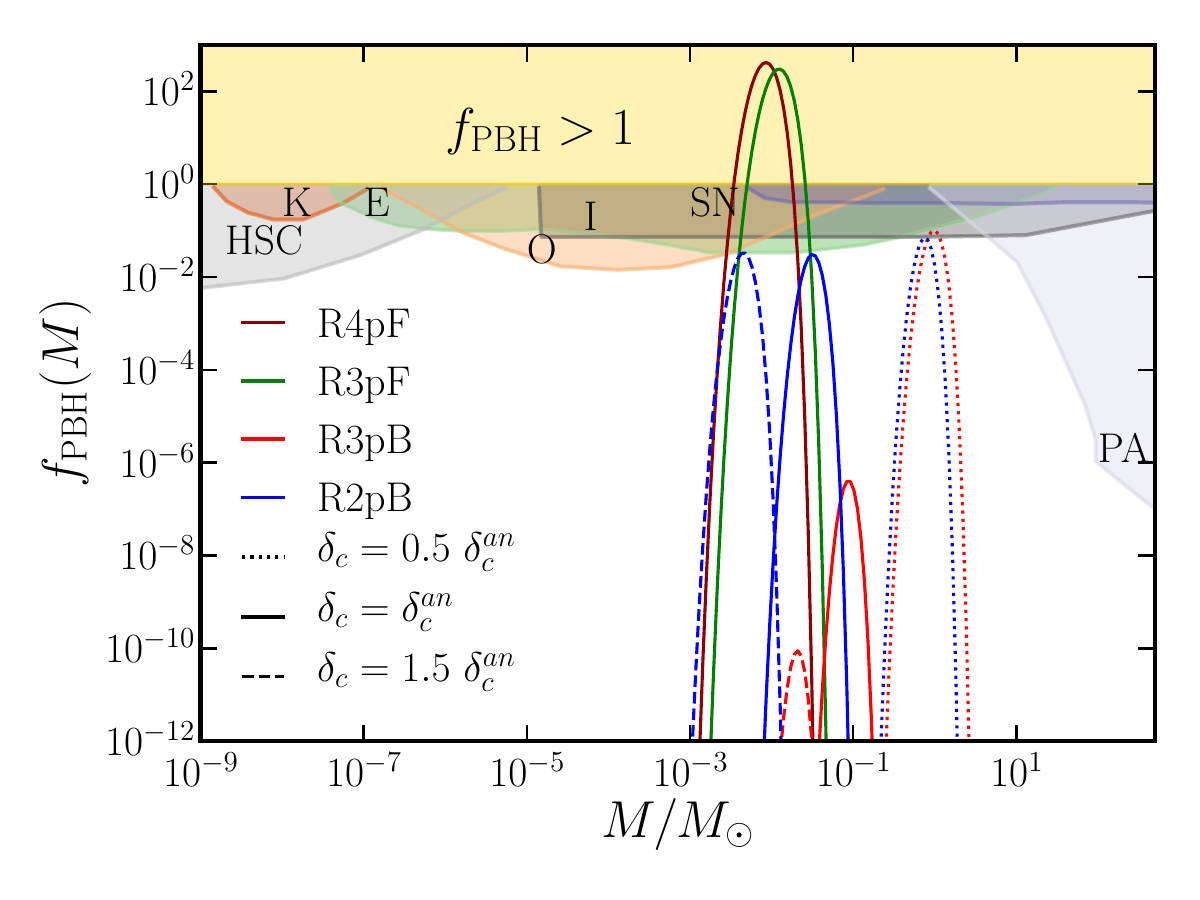}
\caption{The inflationary scalar power spectrum $\cP_\cR(k)$ (on the 
left) and the fraction of PBHs that constitute the energy density of
cold dark matter today~$\fpbh(M)$ (on the right), which correspond 
to the best-fit values of the parameters in the four models are 
plotted.
Clearly, in the case of the models {\tt R4pF} and {\tt R3pF}, the best-fit 
values lead to the overproduction of PBHs.
It is also interesting to note that the best-fit inflationary scalar power 
spectra avoid the current and expected bounds from spectral distortions.}
\label{fig:ogwbest1a}
\end{figure}

In Fig.~\ref{fig:ogwbest1}, we have plotted the spectral density of 
secondary GWs $\ogw(f)$ that correspond to the best-fit values of the 
parameters in the four models {\tt R4pF}, {\tt R3pF}, {\tt R3pB} and 
{\tt R2pB}, which we have considered.
In Fig.~\ref{fig:ogwbest1a}, we have plotted the corresponding inflationary 
scalar power spectra~$\cP_{\cR}(k)$ and fraction of PBHs~$\fpbh(M)$ that 
constitute the energy density of cold dark matter today.
The most important effect in the latter two models is the shift in the set 
of optimal parameters upon the implementation of condition for the formation
of PBHs. 
The latter models prefer comparatively higher values of~$\wre$ and~$\dc$.
As should be clear from the plots of $\fpbh(M)$ in Fig.~\ref{fig:ogwbest1a},
the two original two models, viz. {\tt R4pF} and {\tt R3pF}, though they lead to 
a good fit to the NANOGrav 15-year data, they overproduce PBHs. 
However, as should be evident from the figure, the models {\tt R3pB} and 
{\tt R2pB} are consistent $\fpbh(M)<1$ and the constraints on $\fpbh(M)$,
respectively.
Also, interestingly, we find that, in these models, the higher the $\dc$, 
the smaller is the value of $M$ where the peaks of $\fpbh(M)$ occur.
Lastly, from Fig.~\ref{fig:ogwbest1a}, it should also be clear that all the
models are consistent with the current bounds on the primordial scalar power
spectrum arrived at through the observations of the spectral distortions by 
FIRAS~\cite{1994ApJ...420..439M,2012ApJ...758...76C,2014PhRvL.113f1301J} or 
the future bounds expected from missions such as 
PIXIE~\cite{2011JCAP...07..025K,2021ExA....51.1515C}.


\subsection{Primary GWs and constraints from~$\dneff$}\label{sec:pgw}

We shall now turn to discuss the spectral energy density of primary GWs 
(i.e. the first-order tensor perturbations) that are generated during 
inflation from the quantum vacuum when no sources are present. 
Until now, there is no direct evidence of the primary GWs, but the CMB 
observations lead to an upper bound on their amplitude on large scales.
The constraint on their strength is usually quoted in terms of the 
tensor-to-scalar ratio~$r$, evaluated at or near the CMB pivot scale of 
$k_\ast\simeq 0.05\,\mathrm{Mpc}^{-1}$.
The latest CMB data from BICEP/Keck and Planck suggest that $r \lesssim
0.036$~\cite{BICEP:2021xfz}.

Let $\gamma_k^{\lambda}$ denote the Fourier modes associated with the 
first-order tensor perturbations.
In the spatially flat FLRW universe, they satisfy the following equation
of motion:
\begin{equation}
{\gamma_k^{\lambda}}''+2\,a\,H\,{\gamma_k^\lambda}'
+k^2\,\gamma_k^\lambda=0,\label{eq:eom-pgw}
\end{equation}
which is essentially the same as the equation that describes the Fourier 
modes of the tensor perturbations at the second order [cf. Eq.~\eqref{eq:eom-sgw}],
but without a source.
Consider a slow roll inflationary scenario driven by a single, canonical
scalar field.
In such a case, it is well known that the spectrum of primary GWs produced 
at the end of inflation proves to be nearly scale invariant, with an 
amplitude that is determined by the Hubble scale during inflation.
The perturbations that are generated during inflation need to be evolved 
through the epochs of reheating and radiation domination, in order to arrive 
at the spectral energy density of primary GWs today (in this context, see, 
for example, Refs.~\cite{Giovannini:1998bp,Giovannini:1999bh,Riazuelo:2000fc,
Seto:2003kc,Boyle:2007zx,Stewart:2007fu,Li:2021htg,Artymowski:2017pua,
Caprini:2018mtu,Bettoni:2018pbl,Figueroa:2019paj,Opferkuch:2019zbd,Bernal:2019lpc,Bernal:2020ywq,
Vagnozzi:2020gtf,Mishra:2021wkm,Haque:2021dha,Benetti:2021uea,Giovannini:2022eue}). 
The strength and shape of the spectral density of primary GWs today are sensitive 
to the energy scale of inflation as well as the history of the background evolution 
after inflation.

For simplicity, let us consider inflation to be of the de Sitter form characterized
by the constant Hubble parameter, say, $\HI$\footnote{A priori, such an assumption 
may seem inconsistent with the inflationary model involving an epoch of ultra slow 
roll that is supposed to produce the scalar power spectrum~\eqref{eq:PR}. 
However, it is known that the de Sitter approximation for the scale factor works 
well during the epoch of ultra slow roll.
Moreover, the departure from slow roll only leads to a step in an otherwise nearly
scale-invariant tensor power spectrum (in this regard, see, for instance,
Ref.~\cite{Ragavendra:2023ret}, Fig.~3).}.
In such a situation, as is well known, the scale of inflation can expressed in terms 
of the tensor-to-scalar ratio $r$ and the scalar amplitude $\As$ as follows:~$\HI=\pi\,
\Mpl\,\sqrt{r\, \As/2}$.
Let us first focus on wave numbers that are significantly larger than the wave 
numbers associated with the CMB scales and which re-enter the Hubble radius during 
the epoch of radiation domination.
After they enter the Hubble radius, the energy density associated with the GWs 
behave in the same manner as the energy density of radiation.
Given a scale-invariant spectrum of primary GWs at the end of inflation,  the 
dimensionless spectral energy density of primary GWs today, over wave numbers 
that re-enter the Hubble radius during radiation domination (i.e. over $k<\kre$)
proves to be scale invariant as well, with an amplitude that is given by
\begin{equation}
\ogw(k)\, h^2
= \oR\,h^2\,\f{\HI^2}{12\, \pi^2\, \Mpl^2} 
= 3.5 \times 10^{-17}\, \l(\f{\HI}{10^{-5} \Mpl}\r)^2,
\end{equation}
where~$\oR\,h^2= 4.16\times10^{-5}$ denotes the present-day dimensionless energy 
density of radiation (i.e. including photons and all three species of neutrinos).

If there arises a phase of reheating between inflation and the epoch of radiation
domination, then, evidently, the wave numbers over the range $ \kre < k< \ke$ will 
re-enter the Hubble radius during the epoch of reheating.
Needless to add, the wave number $\kre$ depends on the EoS parameter~$\wre$ and 
the reheating temperature~$\Tre$.
One finds that, over the wave numbers $\kre< k < \ke$, the spectral energy
density of primary GWs today can be expressed as  (see, for instance, 
Ref.~\cite{Haque:2021dha} for detailed calculation)
\begin{align}
\ogw(k)\,h^2 
&\simeq \oR\,h^2\,\f{\HI^2}{12\, \pi^2\, \Mpl^2}\, \f{\mu(\wre)}{\pi}\,
\l(\f{k}{\kre}\r)^{n_{\wre}}\nn\\
&\simeq  3.5 \times 10^{-17}\, \l(\f{\HI}{10^{-5}\, \Mpl}\r)^2\,
\f{\mu(\wre)}{\pi}\,\l(\f{k}{\kre}\r)^{n_{\wre}},\label{eq:ogw-kre-ke}
\end{align}
where the quantity $\mu(\wre)$ and the index~$n_{\wre}$ are given by
\begin{equation}
\mu(\wre)=(1+3\, \wre)^{4/(1+3\,\wre)}\, \Gamma^2\l(\f{5+ 3\,\wre}{2+6\,\wre}\r),
\quad
n_{\wre}=-\f{2\,(1-3\,\wre)}{1+3\,\wre}.
\end{equation}
Note that, while $\mu(\wre) \simeq {\mathcal O}(1)$ for $0\leq \wre \leq 1$, 
$n_{\wre}$ is positive or negative for $\wre>1/3$ or $\wre<1/3$, and 
vanishes $\wre=1/3$.
Moreover, the wave number $\ke$ can be approximately expressed in terms of 
the inflationary energy scale $\HI$ and the reheating parameters $\Tre$ and 
$\wre$ as
\begin{equation}   
\ke = \aend\,\HI \simeq \HI\, \f{T_0}{\Tre}\,
\l(\f{43}{11\, g_{\mathrm{s,re}}}\r)^{1/3}\,
\l({\f{\pi^2\,g_{\mathrm{re}}\,\Tre^4}{90\,\Mpl^2\,\HI^2}}\r)^{1/[3\,(1+\wre)]}.
\label{eq:ke}
\end{equation}

Since, at suitably late times (when the wave numbers of interest are inside 
the Hubble radius), the energy density of GWs behaves in the same manner as 
that of radiation, they can contribute additional relativistic degrees of
freedom to radiation.
These additional number of relativistic degrees of freedom, which is usually 
denoted as $\dneff$, is well constrained by the CMB observations~\cite{Planck:2018vyg}.
This constraint becomes particularly important when $\wre>1/3$ because, in 
such cases, the spectral energy density of primary GWs has a positive 
spectral index for $k>\kre$.
The constraint is given by (see, for instance, 
Refs.~\cite{Jinno:2012xb,Caprini:2018mtu})
\begin{align}
\int_{\kre}^{\ke}\f{\d k}{k}\,\ogw(k)\,h^2
\leq \f{7}{8}\,\l(\f{4}{11}\r)^{4/3}\,\Omega_{\gamma}\,h^2\,\dneff,
\label{eq:deltaneff}
\end{align}
where $\Omega_{\gamma}\,h^2\simeq 2.47\times10^{-5}$ represents the energy 
the density of photons today.
Upon using the form~\eqref{eq:ogw-kre-ke} of $\ogw(k)$, the above condition 
reduces to
\begin{align}
\oR\,h^2\,\f{\HI^2}{12\, \pi^2\, \Mpl^2}\, \frac{\mu(\wre)\,(1+3\,\wre)}{2\,\pi\,
(3\,\wre -1)}\,\l(\f{\ke}{\kre}\r)^\f{6\,\wre -2}{1+3\,\wre}
\leq 5.61\times 10^{-6}\,\dneff.\label{Eq:BBNapprox}
\end{align}
Also, note that the ratio between $\ke$ and $\kre$ can be expressed as 
\begin{equation}
\f{\ke}{\kre}
=\l(\frac{90\,\HI^2\,\Mpl^2}{\pi^2\,g_{\mathrm{re}}}\r)^{(1+3\,\wre)/[6\,(1+\wre)]}\,
\Tre^{-2\,(1+3\,\wre)/[3\,(1+\wre)]}.
\end{equation}
The above three equations can then be combined to arrive at the following 
bound on the reheating temperature $\Tre$:
\begin{equation}
\Tre \geq \l[\f{\oR\, h^2}{5.61\times 10^{-6}\,\dneff}\,
\f{\HI^2}{12\, \pi^2\, \Mpl^2}\, \frac{\mu(\wre)\,(1+3\,\wre)}{2\,\pi\,
(3\,\wre -1)}\,\r]^{\f{3\,(1+\wre)}{4\,(3\,\wre -1)}}\,
\l(\f{90\,\HI^2\,\Mpl^2}{\pi^2\,g_{\mathrm{re}}}\r)^{\f{1}{4}}.
\label{eq:BBNrestriction}
\end{equation}
This condition implies that, for a given $(\Tre,\wre)$, the bound on $\dneff$ 
leads to a restriction on the energy scale of inflation $\HI$ or, equivalently, 
on the tensor-to-scalar ratio~$r$~\cite{Chakraborty:2023ocr}. 
Recall that, in our comparison of secondary GWs with the NANOGrav 15-year data 
in the previous section, we had fixed the reheating temperature to be $\Tre=50\,
\mathrm{MeV}$, just above the standard BBN value of about $4\,\mathrm{MeV}$.
Given $\Tre$, the best-fit values of~$\wre$ from Tab.~\ref{tab:upop} and the
CMB constraint $\dneff < 0.284$~\cite{Planck:2018vyg}, we can calculate the 
value of the tensor-to-scalar ratio~$r$ that is consistent with the 
bound~\eqref{eq:BBNrestriction} above.
In Tab.~\ref{tab:rHmax}, we have listed these maximum allowed values of~$r$, 
for the four models we have considered earlier, viz. {\tt R4pF}, {\tt R3pF},
{\tt R3pB} and {\tt R2pB}. 
\begin{table}[!t]
\begin{center}
\begin{tblr}{|c|c|c|c|c|}
\hline  
\bf{Model} & \SetCell[c=3]{c}{\bf{Maximum allowed value of $r$}} \\
\hline 
{\tt R4pF} & \SetCell[c=3]{c}{$0.036$}\\
\hline
{\tt R3pF} 
& \SetCell[c=3]{c}{$0.036$}\\
\hline
 & {\boldmath $0.5\, \dcan$} &  {\boldmath $\dcan$} & {\boldmath $1.5\,\dcan$}\\
\hline
{\tt R3pB} & $4.2\times10^{-8}$  &$2.7\times10^{-5}$&$0.036$\\
\hline
{\tt R2pB} & $6.8\times10^{-8}$&$2.4\times10^{-3}$&$0.036$\\
\hline
\end{tblr}
\caption{We have listed the maximum allowed value of the tensor-to-scalar
ratio~$r$ that is consistent with the reheating temperature of $\Tre=50\, 
\mathrm{MeV}$, the best-fit values of $\wre$ arrived at on comparison with 
the NANOGrav 15-year data [cf. Tab.~\ref{tab:upop}], and the constraint 
$\dneff < 0.284$ from Planck~\cite{Planck:2018vyg}.
We have listed the values of~$r$ for the four models we discussed earlier, 
viz. {\tt R4pF}, {\tt R3pF}, {\tt R3pB} and {\tt R2pB}.}\label{tab:rHmax}
\end{center}
\end{table}
In Fig.~\ref{fig:ogw1}, we have plotted the dimensionless spectral density of 
primary GWs today corresponding to $r=0.036$, $\tre= 50\,\mathrm{MeV}$ and 
values of $\wre$ for which we had plotted the spectral density of secondary 
GWs generated due to excess scalar power on small scales.
We have plotted the primary spectral energy densities until the frequencies 
that correspond to $\ke$ for each of the parameters, which is determined by the 
relation~\eqref{eq:ke}.


\section{Conclusions and discussion}\label{sec:cd}

Significant amounts of PBHs can be formed in scenarios wherein the inflationary 
scalar power spectrum is considerably amplified on small scales when compared 
to the COBE normalized values on the CMB scales.
In such situations, as the PBHs are formed when the wave numbers re-enter the 
Hubble radius either during the phase of reheating or during the epoch of 
radiation domination, secondary GWs of considerable strengths are inevitably
induced by the excess scalar power on small scales.
In this work, we examined whether the secondary GWs generated in such a scenario
can explain the NANOGrav 15-year data, without the overproduction of solar mass 
PBHs, an issue that has been plaguing some of the recent analyses in this context.
We have listed below the main conclusions of our analysis.
\begin{itemize}
\item 
To begin with, as one would have expected, we find that an extended phase of 
reheating affects the abundance of PBHs as well as the spectral energy 
densities of both the primary and secondary GWs. 
For a given primordial scalar power spectrum, the parameters describing reheating, 
viz. $\tre$ and $\wre$, not only influence the masses of the PBHs formed, but they 
also affect the number of PBHs produced over the corresponding mass range. 
In Fig.~\ref{fig:fpbh}, we illustrated the dependence of $\fpbh(M)$ on the 
parameters~$\kp$, $\tre$ and~$\wre$, assuming fixed values for $A_0$ and $n_0$. 
We find that, for fixed values of $\tre$ and $\wre$, as we increase $\kp$, 
the maximum value of $\fpbh(M)$ behaves as $M^{-2\,\wre/(1+\wre)}$. 
Moreover, for given $\kp$ and $\wf$, as $\tre$ changes, the maximum value 
of $\fpbh(M)$ varies as $M$ when $\wre<1/3$ and as $M^{-1}$ when $\wre>1/3$. 
However, when $\wre$ is varied with the other two parameters fixed, the behavior 
of the maximum value of $\fpbh(M)$ is not linear. 
For a given $\kp$ and $\tre$, the maximum of $\fpbh(M)$ has the lowest value when 
$\wre=1/3$ and it increases on either side as we move away from $\wre$ of $1/3$.
\item 
Secondly, to address the issue of overproduction of PBHs as one attempts to
explain the NANOGrav 15-year data in terms of the secondary GWs generated due
to enhanced scalar power on small scales, we carried out MCMC runs without
and with the constraints on $\fpbh(M)$.
We find that, if we do not include the bounds on $\fpbh(M)$ (as in the case of 
the models {\tt R4pF} and {\tt R3pF}), the best-fit values arrived at upon
comparing the models with the NANOGrav 15-year data indeed leads to the formation 
of excessive number of PBHs  [with $\fpbh(M)>1$].
In order to circumvent this difficulty, we carried out MCMC runs which
ensure that either $\fpbh(M)<1$ (as in the case of the model {\tt R3pB}) or 
remain consistent with the available observational bounds on $\fpbh(M)$ (as 
in the model {\tt R2pB}).
These points should be evident from Fig.~\ref{fig:ogwbest1} wherein we have 
plotted $\fpbh(M)$ for the best-fit values (listed in Tab.\ref{tab:upop})
of the four models.
Importantly, we find that, if we consider the formation of PBHs during a 
prolonged phase of reheating with the EoS parameter lying in the range of 
$0.5$--$0.65$, then, it is possible to remain within the maximum allowed 
abundance of PBHs and simultaneously account for the NANOGrav 15-year data 
with a very strong Bayesian preference (as should be clear from Tab.~\ref{tab:BF}),
when compared to the astrophysical model of SMBHBs. 
\item 
Thirdly, in our analysis, we also accounted for the impact of the
uncertainties in the collapse condition, in particular, the value of
the critical density $\dc$, on the formation of PBHs.
As one can intuitively predict, assuming a lower value of $\dc$ can
lead to an acute overproduction of PBHs, whereas a higher value of
$\dc$ can significantly suppress the number of PBHs formed, thereby 
allowing a larger volume of the parameter space to be compatible with 
the SGWB suggested by the NANOGrav 15-year data.
This is clearly reflected in the Bayesian factor (listed in Tab.~\ref{tab:BF})
for the model~{\tt R2pB}.
Note that the Bayesian factor is considerably higher when $\dc=1.5\,\dcan$
than when $\dc=0.5\,\dcan$.
\item 
Lastly, since the spectral density of primary GWs has a blue tilt for
$\wre>1/3$, it can lead to a violation of the constraint on the number 
of relativistic degrees of freedom during BBN.
We take into account the latest constraints on $\dneff$ to arrive at 
bounds on the tensor-to-scalar ratio~$r$ for the best-fit values of
the parameter~$\wre$ (see Tab.~\ref{tab:rHmax}).
Interestingly, we find that, in the models {\tt R3pB} and {\tt R2pB}, 
the constraint on~$r$ from $\dneff$ proves to be more stringent than 
the bound from the CMB observations~\cite{BICEP:2021xfz}. 
\color{black}
\end{itemize}

In this work, we have remained agnostic about the origin of the primordial
scalar power spectrum.
We have assumed a broken power law form for the spectrum, without specifying 
an inflationary model to generate it.
The form of the spectrum has been motivated by the typical spectra that arise
in single field models of inflation which permit a brief epoch of ultra slow 
roll.
Needless to add, it will be worthwhile to extend our analysis and investigate 
particular inflationary models that generate such spectra.
As non-Gaussianities are expected to be relatively large in scenarios involving 
ultra slow roll inflation~\cite{Ragavendra:2023ret}, constructing such a model 
will not only allow us to explicitly evaluate the non-Gaussianities generated 
in these models, but also study their effects on the generation of both 
PBHs~\cite{Bullock:1996at,Young:2015cyn,Yoo:2019pma} and secondary 
GWs~\cite{Ragavendra:2021qdu}.
Another important aspect that we need to take into account is the effects 
due to a gradual transition from the phase of reheating to the epoch of radiation 
domination. 
In order to consider these effects, we will first need to develop theoretical 
as well as numerical frameworks to study the formation of PBHs in backgrounds 
with a time-dependent EoS parameter.
With the necessary tools at hand, we can study the possibility of such scenarios
to be able to explain the PTA data satisfactorily. 
We are presently working on some of these issues.


\section*{Acknowledgements}

SM and LS wish to thank the Indian Institute of Technology Madras, Chennai, 
India, for support through the Exploratory Research 
Project~RF22230527PHRFER008479. NB wishes to thank his parents for their selfless support that never ceased during this work or any of NB's earlier works. During this work, NB was supported by the institute postdoctoral fellowship (IPDF) of the Indian Institute of Technology Madras, Chennai, India, and later by postdoctoral fellowship of International Centre for Theoretical Physics Asia-Pacific (ICTP-AP), Beijing, China.
NB also thanks Rajeev Kumar Jain and Ranjan Laha for the initial collaborations 
and fruitful discussions on the generation of scalar-induced SGWB during the 
phase of reheating, specifically in the context of the resonant amplification 
of the SGWB. 
MRH wishes to acknowledge support from the Science and Engineering Research 
Board (SERB),  Department of Science and Technology (DST), Government of 
India (GoI), through the National Postdoctoral Fellowship~PDF/2022/002988. 
DM and LS gratefully acknowledge the support received from SERB, DST, GoI, 
through the Core Research Grant~CRG/2020/003664.
LS also wishes to thank the Indo-French Centre for the Promotion of Advanced 
Research for support of the proposal 6704-4 under the Collaborative Scientific 
Research Programme. We also thank our referee in JCAP for his/her fruitful comments, 
which helped us improve the discussions in the manuscript. 
During this work, NB was supported by the institute postdoctoral fellowship (IPDF) of 
the Indian Institute of Technology Madras, Chennai, India, and later by postdoctoral 
fellowship of International Centre for Theoretical Physics Asia-Pacific (ICTP-AP), 
Beijing, China.


\section*{Note added}

During the completion of this work, a preprint appeared on the arXiv [137], which also 
studies the generation of secondary GWs during reheating, induced by primordial scalar 
power spectra that contain either a delta function or a lognormal peak on small scales.
The work also compares the spectral density of GWs with the NANOGrav data.
In the models {\tt R4pF} and {\tt R3pF}, we have also considered primordial
scalar spectra with enhanced power on small scales, but in the form of a 
broken power law.
To ensure that the optimal parameters do not lead to the overproduction of 
PBHs, in our work, we have further carried out runs which {\it simultaneously}\/
calculate $\fpbh(M)$ and scan the parameter space that is restricted to either 
$\fpbh < 1$ (in the model referred to as {\tt R3pB}) or is consistent with 
the available mass-dependent constraints on $\fpbh(M)$ (in the model {\tt R2pB}).
Moreover, in the models {\tt R3pB} and {\tt R2pB}, we have explicitly 
accounted for the uncertainty in~$\dc$.
Further, in the case of the model {\tt R2pB}, using the~{\tt enterprise} mode 
of~{\tt PTArcade}, we have evaluated the Bayesian evidences associated with the optimal 
scenarios for different values of~$\dc$.
We have also considered the primary GWs generated in the scenario and discussed 
the impact of the constraints from $\dneff$ on the tensor-to-scalar ratio or, 
equivalently, the energy scale of inflation.


\bibliographystyle{JHEP}
\bibliography{refs}

\end{document}